\documentclass[11pt]{article}
\usepackage{graphics,cite,amssymb,epsfig,float}
\usepackage[usenames,dvips]{color}
\usepackage{amsmath, mathbbol}

\usepackage{multirow}
\makeatletter

\@addtoreset{equation}{section}
\makeatother

\begin{document}

\begin{titlepage}
\renewcommand{\thefootnote}{\fnsymbol{footnote}}
\setcounter{footnote}{0}

\vspace*{-2cm}
\begin{flushright}
LPT Orsay 11-19 \\
CFTP/11-008 \\
PCCF RI 1102\\

\vspace*{2mm}
\today
\end{flushright}

\begin{center}
\begin{center}
\vspace*{15mm}
{\Large\bf Probing the supersymmetric type III seesaw: 
 LFV  at  low-energies and at the LHC } \\
\vspace{1cm}
{\bf A. Abada$^{a}$, A. J. R. Figueiredo$^{b}$, J. C. Rom\~ao$^{b}$ and 
A. M. Teixeira$^{c}$
}

 \vspace*{.5cm} 
$^{a}$ Laboratoire de Physique Th\'eorique, CNRS -- UMR 8627, \\
Universit\'e de Paris-Sud 1, F-91405 Orsay Cedex, France

\vspace*{.2cm} 
$^{b}$ Centro de F\'{\i}sica Te\'orica de Part\'{\i}culas, 
Instituto Superior T\'ecnico, \\ Av. Rovisco Pais 1, 
1049-001 Lisboa, Portugal

\vspace*{.2cm} 
$^{c}$ Laboratoire de Physique Corpusculaire, CNRS/IN2P3 -- UMR 6533,\\ 
Campus des C\'ezeaux, 24 Av. des Landais, F-63171 Aubi\`ere Cedex, France

\end{center}

\vspace*{10mm}
\begin{abstract}

\vspace*{3mm}
We consider a supersymmetric type III seesaw, where the
additional heavy states are embedded into complete SU(5) representations to
preserve gauge coupling unification. Complying with 
phenomenological and experimental constraints strongly tightens
the viable parameter space of the model. 
In particular, one expects very characteristic signals of
lepton flavour violation both at low-energies and at the LHC, which
offer the possibility of falsifying the model.   

\end{abstract}
\end{center}

\vspace*{3mm}

\end{titlepage}

\renewcommand{\thefootnote}{\arabic{footnote}}
\setcounter{footnote}{0}
\setcounter{page}{2}

\section{Introduction}\label{sec:int}

The seesaw mechanism, in its different realisations,  constitutes one
of the simplest and yet most elegant ways to explain neutrino masses
and mixings. In the minimal realisations of the seesaw, 
the Standard Model (SM) can be extended by the addition
of fermionic singlets (type I seesaw)~\cite{seesaw:I}, scalar triplets 
(type II)~\cite{seesaw:II} or
fermionic triplets (type III)~\cite{seesaw:III}. Although dependent on the 
size of the neutrino Yukawa couplings ($Y^\nu$), these new states are in general
heavy: assuming natural couplings, $Y^\nu \sim 1$, 
their masses can be close to the 
Grand Unified Theory (GUT) scale, $\mathcal{O}(10^{16}\text{ GeV})$. 

If these states are indeed at the origin of neutrino mass generation, it is
important to investigate which seesaw realisation (or combination thereof) is
at work. Indeed, if the mass of the mediators is such that production at present
colliders is possible (in this case $Y^\nu \sim 10^{-6}$), 
then one can devise strategies for their direct searches. 
On the contrary, if they are  very heavy, then they cannot be directly
probed, and their indirect signatures in low-energy observables
(typically 
via higher order
corrections) will be extremely suppressed. 

Other than the mechanism of neutrino mass generation, there are several
reasons - theoretical issues and observational problems - 
motivating the extension (or embedding) of the SM into a larger
framework. 
Supersymmetry (SUSY) is a well
motivated solution for the hierarchy problem that also offers an
elegant solution for the non-baryonic dark matter (DM) problem of 
the Universe~\cite{Jungman:1995df, Bertone:2004pz, WMAP}.  If the 
Large Hadron Collider (LHC)
indeed finds signatures of SUSY, it is then extremely appealing to
consider the embedding of a seesaw mechanism
into a supersymmetric framework (the so-called SUSY seesaw). 

Supersymmetric seesaws lead to a number of possible signatures in the
neutral and charged lepton sectors, both at
low and high energies. Among low-energy observables, the most
striking SUSY seesaw impact is perhaps the possibility of having
charged lepton flavour violating (LFV) transitions. Indeed,
one can have sizable contributions to
radiative decays ($\ell_i \to \ell_j \gamma$), three-body decays ($\ell_i \to 3
\ell_j$)  and  $\mu-e$ transitions in heavy 
nuclei, well
within reach of current and/or future dedicated facilities~\cite{Hisano:1995cp, Hisano:1995nq, Hisano:1998fj,
  Buchmuller:1999gd, Kuno:1999jp, Casas:2001sr, Lavignac:2001vp,
  Bi:2001tb, Ellis:2002fe,Deppisch:2002vz, Fukuyama:2003hn,
  Brignole:2004ah,  Masiero:2004js, Fukuyama:2005bh, Petcov:2005jh,
  Arganda:2005ji, Deppisch:2005rv, Yaguna:2005qn, Calibbi:2006nq,
  Antusch:2006vw, Hirsch:2008dy, Arganda:2007jw, Arganda:2008jj}. 
At high-energy colliders, such as the LHC, several observables may 
reflect an underlying SUSY seesaw. Let us begin by noticing that
if some components of the seesaw mediators 
are not singlets under the SM gauge group 
(which is the case in type II and III seesaws), the latter can leave an
imprint on the SUSY spectrum, since they can modify the supersymmetric
$\beta-$ functions governing the evolution of the gauge couplings and
soft-SUSY breaking parameters. At the LHC, SUSY seesaws can also give
rise to several LFV signals: firstly, one can have sizable widths for LFV decay
processes like $\chi_2^0 \to \chi_1^0\ \ell_i^\pm\, \ell_j^\mp
$~\cite{Arkanihamed:1996au, Hinchliffe:2000np,
Carvalho:2002jg, Hirsch:2008dy, Carquin:2008gv}; secondly, one can
have observable flavoured slepton mass splittings (MS),   
$\Delta  m_{\tilde{\ell}}/m_{\tilde{\ell}} \ ({\tilde e_{_L}}, 
{\tilde \mu_{_L}})$ and $\Delta  m_{\tilde{\ell}}/m_{\tilde{\ell}} 
\ ({\tilde \mu_{_L}}, {\tilde \tau_{_2}})$. 
These splittings can be identified
since, under certain conditions, one can effectively reconstruct
slepton masses via observables such as the kinematic end-point of the
invariant mass distribution of the leptons coming from the cascade
decays $\chi_2^0 \rightarrow \tilde{\ell}^{\pm} \ell^{\mp} \rightarrow \chi_1^0\ 
 \ell^{\pm}\ell^{\mp}$. If the slepton in the decay chain is
real (on-shell), the di-lepton invariant mass spectrum has a kinematical edge
that might then be measured with a very high precision (up to 0.1
\%)~\cite{Paige:1996nx, Hinchliffe:1996iu, Bachacou:1999zb}.  Together
with data arising from other observables, this information allows to
reconstruct the slepton masses \cite{Paige:1996nx, Hinchliffe:1996iu,
Bachacou:1999zb, Ball:2007zza,ATLAS} and hence study the slepton mass splittings. 
 Finally, one can observe
multiple edges in di-lepton invariant mass distributions from $\chi_2^0 \to
\chi_1^0\  \ell_i^{\pm} \ell_i^{\mp}$, arising from the exchange of a
different flavour slepton $\tilde \ell_j$ (in addition to the left- and
right-handed sleptons, $\tilde \ell_{_{L,R}}^i$). Under the assumption of a
seesaw as the unique underlying source of flavour violation in the
leptonic sector (for instance assuming that SUSY breaking is due to
flavour blind interactions), 
then all the above observables, both at high and low
energies, will be strongly correlated. 

Each seesaw realisation will have a distinct impact on the latter
observables. It is thus mandatory to conduct an
exhaustive study of the many possible experimental signatures, in order to test
the seesaw hypothesis, either excluding or substantiating it, and in
the latter case, devising a strategy to disentangle among the
different seesaw realisations. 

In a previous work~\cite{Abada:2010kj} we have studied the impact of a
type I seesaw, embedded into the constrained minimal supersymmetric extension of the 
SM (cMSSM), in what concerns lepton flavour violation both at low-energies
and at the LHC. 
Here we extend the analysis to the type III SUSY seesaw.  
In this case, and in order to accommodate neutrino masses and mixings, 
one adds (at least two) fermionic SU(2) triplets to
the SM particle field content~\cite{Foot:1988aq}, 
as well as the corresponding superpartners.  
If one extends the usual MSSM by just the superfields responsible
for neutrino masses and mixings, one would destroy the nice feature of
gauge coupling unification.
This problem is easily circumvented  by embedding the new
states in complete SU(5) representations, 
{\bf 24}-plets in the case of a type III
seesaw~\cite{Buckley:2006nv}. Note that in addition to the SU(2) triplet, 
the {\bf 24}-plet contains a singlet state which also
contributes to neutrino dynamics, so that in this case one actually has 
a mixture between type I and type III seesaws.

Our study shows that if a type III seesaw is indeed the unique source
of  neutrino 
masses and leptonic mixings, and is realised within an otherwise
flavour conserving 
SUSY extension of the SM (specifically the cMSSM), one then expects low-energy LFV
observables within future sensitivity reach, as well as interesting slepton
phenomena at the LHC.  
After having identified regions in the cMSSM parameter space, where
the slepton masses could in principle be reconstructed from the
kinematical edges of di-lepton mass 
distributions (i.e.  $\chi_2^0 \to  \chi_1^0\, \ell_i^\pm\, \ell^\mp_i$ 
can occur, and with a non-negligible
number of events), we study the different slepton mass splittings,
exploring the implications for LFV decays. From the comparison
of the predictions for the two sets of observables (high- and 
low-energy) with the current experimental bounds and future sensitivities,
one can either derive information about the otherwise unreachable
seesaw parameters, or disfavour the type III SUSY seesaw as being the unique
source of lepton flavour violation.

The paper is organised as follows:  in Section~\ref{sec:SUSYtypeIII}
we define the model, providing a brief overview on the implementation
of a type III seesaw in the cMSSM. In 
Section~\ref{sec:lfv} we discuss the implications of this mechanism   
for low- and high-energy LFV observables.  Our results are presented in
Section~\ref{sec:results} where we study the different high- and
low-energy observables in the seesaw case. This will also allow to
draw some conclusions on the viability of a supersymmetric 
type III seesaw as the
underlying mechanism of LFV. Further discussion is presented in the
concluding Section~\ref{sec:conclusions}.

\section{Type III SUSY seesaw}\label{sec:SUSYtypeIII}
Under the hypothesis that neutrinos are Majorana particles, the
smallness of their masses, as well as their mixings, can be explained
via seesaw-like mechanisms due to the exchange of heavy states: 
fermionic singlets (type I), scalar triplets (type II) 
or fermionic triplets (type III). 
The three possible seesaw realisations can be easily
embedded in the framework of supersymmetric models. However, if 
SUSY's appealing feature of gauge coupling (and gaugino mass) unification 
is to be preserved,  the new particles present below the 
Grand Unified scale must belong to complete GUT
representations. 
Under the assumption of an SU(5) gauge group, generating a neutrino mass
matrix with at least two non-zero eigenvalues\footnote{Rank $\geq 2$
  neutrino mass matrices can also be obtained with a truly minimal
  heavy field content, via the inclusion of non-renormalisable
  operators in the superpotential (see, for
  example,~\cite{Bajc:2006ia,Biggio:2010me}).} 
requires the following multiplet content: two copies of $\pmb{1}$ or 
$\pmb{24}$ (type I and III, respectively) or $\pmb{\overline{15}}+\pmb{15}$
(type II).
The addition of the non-singlet fields (i.e. the $\pmb{15}$- 
and the $\pmb{24}$-plets) has an important effect on the
evolution of  several fundamental parameters, especially on the $\beta$-functions
for gauge and Yukawa couplings, as well as 
on the Renormalisation Group (RG) running of
mass-terms above the seesaw 
scale. At low-energies (electroweak scale) this translates into
changes in the SUSY spectrum, leading to scenarios that can
be significantly different from a minimal supergravity (mSUGRA) inspired 
(or minimal type I SUSY seesaw) scenario. In turn, this will have
consequences concerning 
flavour observables and cosmological quantities like  the dark
matter relic density. 

We consider in this study a generic framework where three families of
triplet fermions, $\Sigma_i$ (as well as their superpartners) are
added to the MSSM particle 
content~\cite{Esteves:2010ff}. 
Each one is embedded into a $\pmb{24}$-plet\footnote{Among the
  representations of lower dimension, only the
  $\pmb{24}$ does indeed contain a singlet hypercharge field.}, 
that decomposes under the SM
gauge group,
SU(3)$\times$SU(2)$\times$U(1), as
\begin{eqnarray}\label{eq:24decomposition}
\pmb{24}\,&=&
\,(8,1,0)\,+\,(3,2,-5/6)+\,(3^*,2,5/6)\,+\,(1,3,0)\,+\,(1,1,0)\, 
\nonumber\\
&=&\,\widehat G_M + \widehat X_M + \widehat{\overline X}_M + \widehat W_M +
\widehat B_M \,.
\end{eqnarray}
The fermionic components of the last two terms in the above
decomposition ($\widehat W_M$ and $\widehat B_M$) have exactly the
same quantum numbers of a fermionic triplet ($\Sigma$) and of a singlet
right-handed neutrino ($\nu_R$). It is then clear that if embedded
into an SU(5) framework, the realisation of the type III seesaw will
in general produce a mixture of type I and type III mechanisms.

In the unbroken SU(5) phase, the superpotential is given by
\begin{equation}\label{eq:SU(5):W}
\mathcal{W}^{\text{SU(5)}}\,=\,
\sqrt{2}\,\pmb{\bar 5}_{M_i}\,Y^5_{ij}\,\pmb{10}_{M_j}\,\pmb{\bar 5}_H -
\frac{1}{4}\pmb{10}_{M_i}\,Y^{10}_{ij}\,\pmb{10}_{M_j}\,\pmb{5}_H + 
\pmb{5}_H\,\pmb{24}_{M_i}\,Y^{N}_{ij}\,\pmb{\bar 5}_{M_j} +
\frac{1}{4}\pmb{24}_{M_i}\,{M_{24}}_{ij}\,\pmb{24}_{M_j}\,,
\end{equation}
with $i,j$ denoting generation (flavour) indices, and where 
we have not included the terms specifying the Higgs sector
responsible for the breaking of SU(5). The Majorana mass term in 
Eq.~(\ref{eq:SU(5):W}) is gauge invariant due to having the
triplet superfields in the adjoint SU(2) representation. 
In the broken phase, in addition to the usual MSSM terms, the
superpotential is given by: 
\begin{eqnarray}\label{eq:brokenSU(5):W}
\mathcal{W}\,&=&\,\mathcal{W}^{\text{MSSM}} + 
\widehat H_2 \left( \widehat W_M\, Y_M - \sqrt{\frac{3}{10}}\,\widehat B_M\, Y_B
\right)\,\widehat L + \widehat H_2\,\widehat{\overline X}_M\,Y_X \,
\widehat D^c + \nonumber
\\
&&+\frac{1}{2}\, \widehat B_M\,M_B\,\widehat B_M+\frac{1}{2}\, 
\widehat G_M\,M_G\,\widehat G_M+
\frac{1}{2}\, \widehat W_M\,M_W\,\widehat W_M+ \widehat X_M\,M_X\,
\widehat{\overline X}_M\,.
\end{eqnarray}
After the heavy fields are integrated out, and at lowest order in the
expansion in $(v_2\, Y_{B,W}/M_{B,W})^n$ ($v_2$ being the vacuum expectation 
value of $H_2^0$), one obtains the light
neutrino mass matrix:
\begin{equation}\label{eq:seesaw.BW}
m_\nu\, \approx \,-{v_2^2}\,\left(\frac{3}{10}\, Y_B^T\,M_B^{-1}\,Y_B
\,+\,\frac{1}{2}\, Y_W^T\,M_W^{-1}\,Y_W \right),
\end{equation}
where we have again omitted flavour indices. From the above formula, it is
clear that we are indeed in the presence of a mixed type I and III
seesaw, with contributions to $m_\nu$ arising from both the singlets
($\propto Y_B$) and SU(2) triplets ($\propto Y_W$).
The model is further specified by $\mathcal{W}^{\text{MSSM}}$ and by
the soft-SUSY breaking Lagrangian. Concerning the latter, we will furthermore 
assume a cMSSM framework,
where mSUGRA-inspired universality conditions for the
soft-breaking SUSY parameters are imposed at some very high-energy
scale, which we take to be $M_\text{GUT}$. The MSSM part of the model
is then defined by the usual 4 continuous parameters (the universal
gaugino and scalar soft-breaking masses $M_{1/2}$ and $m_0$, the
universal trilinear coupling $A_0$ and the ratio of the Higgs vacuum
expectation values, $\tan \beta=v_2/v_1$) and the sign of the bilinear
$\mu$-term in $\mathcal{W}^{\text{MSSM}}$, sign($\mu$). 

One can further impose
additional GUT scale SU(5)-motivated boundary conditions for the Yukawa
couplings and Majorana masses appearing in
Eq.~(\ref{eq:brokenSU(5):W}): $Y_B=Y_W=Y_X$ and
$M_B=M_W=M_G=M_X$. Although the above parameters do run between the
GUT scale and their corresponding decoupling scales, one has, to a very
good approximation, that  $Y_B \simeq Y_W$ and $M_B \simeq M_W$ as the
heavy states decouple. At the seesaw scale (which we define to be 
$\approx M_B\simeq M_W$) $m_\nu$ is approximately given by
\begin{equation}\label{eq:seesaw}
m_\nu\, \approx \,-v_2^2\,\frac{4}{5}\,
Y^{{\nu^T}} M_N^{-1}\,Y^{\nu}\,,
\end{equation}
where we have again used the simplifying notation $Y_B=Y_W=Y_N=Y^{\nu}$,  
$M_B=M_W=M_N$. 

Up to an overall factor ($4/5$), one can still use 
the Casas-Ibarra parametrization~\cite{Casas:2001sr} for 
the neutrino Yukawa couplings at the seesaw scale $M_N$,
\begin{equation}\label{eq:seesaw:casas}
Y^\nu v_2=\, i \sqrt{M^\text{diag}_N}\, R \,
\sqrt{m^\text{diag}_\nu}\,  {U_\text{MNS}}^{\dagger}\,.
\end{equation}
In the above $R$ is a complex orthogonal $3 \times 3$ matrix that
encodes the possible mixings involving the heavy neutral states, in
addition to those of the low-energy sector
(i.e. $U_{\text{MNS}}$), and which is parametrized in terms of three 
complex angles $\theta_i$ $(i=1,2,3)$. 

As extensively discussed in~\cite{Esteves:2010ff}, the
$\beta$-functions of the gauge couplings, as well as the running for
soft gaugino and scalar masses, are strongly affected in type III seesaw
models. In fact, RGE effects are behind the relatively small interval
for $M_N$ in a type III SUSY seesaw. Assuming that the triplet masses
are degenerate ($M_{N_i} = M_{24}$),  the interval is bounded from
above, $M_{24} \lesssim 9 \times 10^{14}$ GeV, 
to comply with the atmospheric neutrino mass difference.  
On the other hand, for triplet
masses below $10^{13}$ GeV, the running is such that one encounters
Landau poles for the gauge couplings at the GUT scale, while
tachyonic sfermions (especially the lighter stau and stop) can also
arise for smaller values of the soft-SUSY breaking parameters. 

As clear from the above discussion, the new distinctive features of a 
type III seesaw will likely be manifest in many phenomena. 
In what follows we discuss the new contributions of the type III SUSY seesaw
for low-energy lepton flavour violation (e.g. to radiative decays such as 
$\mu\to e \gamma$), as well as for LFV at the LHC: in particular, 
we focus on the study of slepton mass splittings to probe deviations 
from the cMSSM and possibly derive information about the SUSY seesaw
parameters.

\section{Lepton flavour violation in a type III SUSY seesaw}\label{sec:lfv}

As for the case of a type I SUSY seesaw, the non-trivial flavour
structure of $Y^\nu$ at the GUT scale will induce (through the running
from $M_\text{GUT}$ down to the seesaw scale) flavour mixing in the
otherwise approximately flavour conserving soft-SUSY breaking terms.
In particular, there will be radiatively induced flavour mixing in the slepton
mass matrices, manifest in the $LL$ and $LR$ blocks of the
$6\times 6$ slepton mass matrix; an analytical estimation using the
leading order (LLog) approximation leads to the following corrections
to the slepton mass terms~\cite{Esteves:2010ff}: 
\begin{align}\label{eq:LFV:LLog}
(\Delta m_{\tilde{L}}^2)_{_{ij}}&\,=
\,
- \frac{9}{5}\,\frac{1}{8\, \pi^2}\,  (3\, m_0^2+ A_0^2)\, ({Y^{\nu}}^\dagger\, 
L\, Y^{\nu})_{ij} 
\,,
\nonumber \\
(\Delta A_l)_{_{ij}}&\,=
\,
- \frac{9}{5}\,
\frac{3}{16 \,\pi^2}\, A_0\, Y^l_{ij}\, ({Y^{\nu}}^\dagger\, L\, Y^{\nu})_{ij}
\,,
\nonumber \\
(\Delta m_{\tilde{E}}^2)_{_{ij}}&\,\simeq
\,
0\,\,;\, L_{kl}\, \equiv \,\log \left( \frac{M_X}{M_{N_k}}\right) \,
\delta_{kl}\,.
\end{align}
When compared to the type I SUSY seesaw, the most important difference
corresponds to a change in the overall factor (multiplying the 
$({Y^{\nu}}^\dagger\, L\, Y^{\nu})_{ij}$ term). The above sources of
flavour mixing will have an impact regarding lepton flavour
non-universality and lepton flavour violation in the charged slepton sector,
potentially inducing sizable contributions to high- and low-energy LFV
observables, as we proceed to discuss.

\bigskip
As mentioned in the Introduction, several LFV signals can be
observable at the LHC, in strict relation with the ${\chi}^0_2 \to {\chi}^0_1\,
\ell^{\pm}\,{\ell^{\mp}}$ decay chains. 
As discussed in~\cite{Paige:1996nx,
Hinchliffe:1996iu,Bachacou:1999zb,Ball:2007zza,ATLAS}, in 
scenarios where the
${\chi}^0_2$ is sufficiently heavy to decay via a real (on-shell)
slepton, the process ${\chi}^0_2 \to
{\chi}^0_1\, \ell^{\pm}\,{\ell^{\mp}}$ is greatly enhanced while
providing a very distinctive signal: same-flavour
opposite-charged leptons with missing energy. 
The $\chi_2^0 \to \chi_1^0\,  \ell^\pm\, \ell^\mp$ decay chain 
thus offers a golden laboratory to study LFV at the LHC, via the
following observables:\\
\noindent
(i) sizable widths for  LFV decay processes like
$\chi_2^0 \to \chi_1^0\ \ell_i^\pm\, \ell_j^\mp$~\cite{Arkanihamed:1996au,
Hinchliffe:2000np, Carvalho:2002jg, Hirsch:2008dy, Carquin:2008gv};

\noindent
(ii) multiple edges in di-lepton invariant mass
distributions $\chi_2^0 \to \chi_1^0 \ \ell_i^{\pm} \ell_i^{\mp}$,
arising from the 
exchange of a different flavour slepton $\tilde l_j$ 
(in addition to the left- and right-handed sleptons, $\tilde l_{_{L,R}}^i$);

\noindent
(iii) flavoured slepton mass splittings.

In order to optimise the reconstruction
of the leptons' momentum 
(which is expected to be easy, accounting for
smearing effects in $\tau$'s) and, in addition, extract indirect
information on the mass spectrum of the involved sparticles, 
the SUSY spectrum must comply with the requirements of a so-called 
``standard window'': \\
\noindent
(a) the spectrum is such that the decay chain 
$ \chi_2^0\to \tilde  \ell \  \ell \to \chi_1^0\  \ell \ \ell$, 
with intermediate real sleptons, is allowed;

\noindent
(b) it is possible to have sufficiently hard outgoing 
leptons: $m_{\chi_2^0}-m_{\tilde  \ell_{_L}, \tilde \tau_{_2}} > 10$ GeV. 

\noindent In this case, the di-lepton invariant mass spectrum 
has a kinematical edge that
might be measured with a very high precision (up to 0.1
\%)~\cite{Paige:1996nx, Hinchliffe:1996iu, Bachacou:1999zb}. 
Together with data arising from other observables, this information 
allows to reconstruct the slepton masses \cite{Paige:1996nx,
  Hinchliffe:1996iu, 
Bachacou:1999zb,Ball:2007zza,ATLAS}, and hence probe slepton mass
universality or test LFV in the slepton sector. 
In particular, the relative 
slepton mass splittings, which are defined as
\begin{equation}\label{eq:MS:def}
\frac{\Delta m_{\tilde \ell}}{m_{\tilde \ell}} (\tilde \ell_i, \tilde
  \ell_j) \, = \, 
\frac{|m_{\tilde \ell_i}-m_{\tilde \ell_j}|}{<m_{\tilde
    \ell_{i,j}}>}\,,
\end{equation}
can be inferred from the kinematical edges with a sensitivity of
$\mathcal{O}(0.1\%)$~\cite{Allanach:2008ib} for $\tilde e_{_L} - \tilde
\mu_{_L}$ and $\mathcal{O}(1\%)$ for $\tilde \mu_{_L} - \tilde \tau_{_2}$.

Even in the absence of a seesaw mechanism, it is important to recall that 
universality between the third and first two slepton generations
is broken due to $LR$ mixing and to 
RGE effects proportional to the third generation lepton
Yukawa coupling. 
However, 
in the presence of flavour violation (as induced by the SUSY 
seesaw, see Eqs.~(\ref{eq:LFV:LLog})), the mass differences between
left-handed selectrons, smuons and staus can be potentially
augmented. Similar to 
the case of
a type I
seesaw~\cite{Abada:2010kj}, the relative mass splitting between
left-handed sleptons is approximately given by 
\begin{equation}\label{eq:MS:ij:seesaw}
\frac{\Delta m_{\tilde \ell}}{m_{\tilde \ell}} (\tilde \ell_i, \tilde \ell_j)
\approx 
\frac{|(\Delta m_{\tilde L}^2)_{_{ij}}|}{ m_{\tilde \ell}^2 }
\end{equation}
where we have neglected $LR$ mixing effects, as well as RGE
contributions proportional to the charged lepton Yukawa coupling. 
In the  $R=1$ seesaw limit, where all flavour violation in $Y^\nu$
stems from the $U_\text{MNS}$ (see Eq.~(\ref{eq:seesaw:casas})), 
and assuming that the large flavour violating entries 
involving the second and third generation constitute the dominant
source of mixing (and are thus at the origin of the slepton mass
differences), one can further relate the $\tilde e_{_L}-\tilde \mu_{_L}$
and the $\tilde \mu_{_L}-\tilde \tau_{_2}$ mass
differences~\cite{Abada:2010kj}:
\begin{equation}\label{eq:MS:mutau:emu:R1seesaw}
\frac{\Delta m_{\tilde \ell}}{m_{\tilde \ell}} (\tilde e_{_L}, \tilde
\mu_{_L}) \, \approx \, \frac{1}{2}\,
\frac{\Delta m_{\tilde \ell}}{m_{\tilde \ell}} (\tilde \mu_{_L}, \tilde
\tau_{_2})\,. 
\end{equation}
As discussed in~\cite{Abada:2010kj}, in the framework of a
type I seesaw, the slepton mass differences can be sufficiently
large as to be within the reach of LHC sensitivity. 

Before proceeding, let us briefly notice that, 
depending on the amount of flavour violation, one can be led to 
regimes where two 
non-degenerate mass eigenstates have almost identical flavour content
(maximal flavour mixing). To correctly
interpret a mass splitting between sleptons with quasi-degenerate
flavour content, one has to introduce an ``effective'' mass
\begin{equation}\label{eq:effective:mass}
m^{\text{(eff)}}_i \equiv \sum_{X = \tilde{\tau}_{_2}  ,\,\tilde{\mu}_{_L}  ,\,
\tilde{e}_{_L}} m_{\tilde{l}_X} \left( | R^{\tilde{l}}_{X i_L} |^2 + |
R^{\tilde{l}}_{X i_R} |^2 \right)\ , 
\end{equation}
where $R^{\tilde{l}}$ is the matrix that diagonalizes the $6\times 6$ slepton mass 
matrix. The effective mass splittings are then defined as 
\begin{equation}\label{eq:effective:MS}
\left( \frac{\Delta m}{m} \right)^{\text{(eff)}}(\tilde{\ell}_i,
\tilde{\ell}_j)\, \equiv \,\frac{\,2 \,| m^{\text{(eff)}}_i - m^{\text{(eff)}}_j
  |}{m^{\text{(eff)}}_i + m^{\text{(eff)}}_j} \,. 
\end{equation}

\bigskip
The seesaw-generated flavour violating entries of
Eqs.~(\ref{eq:LFV:LLog}) will also give rise to low-energy LFV
phenomena, such as radiative $\ell_i \to \ell_j \gamma$ decays, which
are induced by 1-loop diagrams via the exchange of gauginos and sleptons. 
These can be described
by the effective Lagrangian~\cite{Hisano:1995cp},
\begin{equation}
\label{eq:LagMuEG}
\mathcal{L}_\text{eff}\, =\, e \,\frac{m_{\ell_i}}{2}\, \bar{\ell}_i \,
\sigma_{\mu \nu} F^{\mu \nu} 
(A_L^{ij} \,P_L + A_R^{ij}\, P_R)\, \ell_j + \text{h.c.} \,,
\end{equation}
where $P_{L,R} = \frac{1}{2}(1 \mp \gamma_5)$ are the usual chirality
projectors and  the couplings $A_L$ and $A_R$ arise from loops involving 
left- and right-handed sleptons, respectively.  Using
Eq.~(\ref{eq:LagMuEG}), the branching ratio 
$\ell_i \to \ell_j \gamma$ is given by
\begin{equation} 
\label{brLLG}
\text{BR}(\ell_i \to \ell_j \gamma) = \frac{48 \pi^3 \alpha}{G_F^2} 
\left( |A_L^{ij}|^2 + |A_R^{ij}|^2 \right) \text{BR}(\ell_i \to \ell_j \nu_i
\bar{\nu}_j) \,. 
\end{equation}
where $G_F$ is the Fermi constant and $\alpha$ is the electromagnetic
coupling constant.  In our numerical calculation we use the exact
expressions for $A_L$ and $A_R$\footnote{The exact formulae for the
branching ratios of the radiative LFV decays, as used in our 
numerical computation,
can be found, for example,  
in~\cite{Raidal:2008jk}.}.  However, for an easier understanding of
the numerical results, we note that the relations between these
couplings and the slepton soft-breaking masses are approximately given by
\begin{equation} 
\label{eq:BR:MIA:LL}
|A_L^{ij}| \sim \frac{|(\Delta m_L^2)_{_{ij}}|\, 
\tan\beta}{m_\text{SUSY}^4} \simeq 
\left|\frac{9}{5}\,\frac{\tan\beta}{8\, \pi^2}\,  \frac{(3\, m_0^2+
    A_0^2)}{m_\text{SUSY}^4}\, ({Y^{\nu}}^\dagger\, L\,
  Y^{\nu})_{ij}\right| , \quad 
A_R^{ij} 
\sim \frac{(\Delta m_{E}^2)_{_{ij}}\, \tan\beta}{m_\text{SUSY}^4} 
\simeq 0\,, 
\end{equation}
where $m_\text{SUSY}$ denotes
a generic (average) SUSY mass and where we have used Eqs.~(\ref{eq:LFV:LLog}).

It is important to notice here that when compared to 
other seesaws, and for the same cMSSM parameters, the sparticle spectrum is
lighter. Together with the larger Yukawa couplings (a 
consequence of the larger seesaw scale), the type III  seesaw
typically leads to larger LFV observables than in either type I 
or II~\cite{Esteves:2010ff}.

Equally interesting LFV observables are 
$\mu-e$ conversions in heavy nuclei, as they offer challenging experimental
prospects: the possibility of improving 
experimental sensitivities to values as
low as $ \sim10^{-18}$ renders this observable an extremely powerful
probe of LFV in the muon-electron sector. In the limit of  
photon-penguin dominance, the conversion rate CR($\mu-e$,\ N) 
and BR($\mu \to e \gamma$) are strongly correlated, 
since both observables are sensitive to the same leptonic mixing
parameters~\cite{Arganda:2007jw}. 

In the following section we  numerically analyse the above discussed points.

\section{Numerical results and discussion}\label{sec:results}

For the numerical computation, we have used the public code {\sc
SPheno} (v3.beta.51)~\cite{Porod:2003um} to carry out the numerical
integration of the RGEs. The RGEs of the SU(5) type III SUSY seesaw were
calculated at 2-loop level in~\cite{Esteves:2010ff}, 
using the public code {\sc SARAH}\cite{Staub:2008uz}. 
{\sc SPheno} further computes the sparticle and Higgs spectrum, as
well as the various low-energy LFV observables.
The dark matter relic density is evaluated through a link to
{\sc{micrOMEGAs}} v2.2~\cite{Belanger:2008sj}.

Regarding low-energy neutrino data, 
current (best-fit) analyses favour the
following intervals for the mixing angles~\cite{GonzalezGarcia:2010er}
\begin{align}\label{eq:mixingangles:data}
& 
\theta_{12}\,=\ (34.4\pm 1.0)^\circ , 
\quad 
\theta_{23}\,=\, (42.8\,  ^{+ 4.7}_{-2.9} )^\circ ,
\quad 
\theta_{13}\,=\, (5.6\,  ^{+ 3.0}_{-2.7} )^\circ \,(\leq 12.5^\circ),
\end{align}
while for the mass-squared differences one has
\begin{align}\label{eq:lightmasses:data}
& 
\Delta\, m^2_\text{21} \,=\,(7.6\, \pm 0.2)\,\times 10^{-5}\,\,\text{eV}^2\,,
\quad 
\Delta \, m^2_\text{31} \,=\left\{ \begin{array}{l} \,(-2.36\, \pm \
    0.11)\,\times 10^{-3}\,\,\text{eV}^2\,\\  
\,(+2.46\, \pm \ 0.12)\,\times 10^{-3}\,\,\text{eV}^2\ 
 \end{array}\right. \,,
\end{align}
where the two ranges for $\Delta \, m^2_\text{31}$ correspond to
 inverted and normal neutrino spectrum.  In Table~\ref{table:LFV:bounds} we 
summarise the current bounds 
and the future sensitivities of dedicated experimental
facilities, for the low-energy LFV observables considered in our
numerical discussion.\\
\begin{table}[h!]
\begin{center}
\begin{tabular}{|l|c r|c r|}
\hline
LFV process & Present bound & & Future sensitivity & \\
\hline
BR($\mu \to e \gamma$) & $1.2 \times 10^{-11}$&
\cite{PDG}&$10^{-13} $ & \cite{Kiselev:2009zz} \\ 
BR($\tau \to \mu \gamma$) & $4.5 \times 10^{-8}$&
\cite{Hayasaka:2007vc}&$ 10^{-9}$ & \cite{Bona:2007qt} \\ 
\hline
CR($\mu-e$, Ti) & $4.3 \times 10^{-12}$ & \cite{PDG}&
${\cal{O}}(10^{-16})$ (${\cal{O}}(10^{-18})$) &
\cite{Glenzinski:2010zz}~(\cite{Cui:2009zz}) \\ 
\hline
\end{tabular}
\end{center}
\caption{Present bounds and future sensitivities for several LFV
  observables.  }
\label{table:LFV:bounds}
\end{table}

\bigskip
In the first part of the analysis we  assume a 
degenerate spectrum for the three families of
triplet fermions. Moreover, we consider the
conservative limit\footnote{In general, the limit $R=1$
translates into a ``conservative'' limit for flavour violation:
apart from possible cancellations, and for a fixed
seesaw scale, this limit typically provides
a lower bound for the amount of generated LFV.} in which  
flavour violation solely arises from the $U_\text{MNS}$ leptonic mixing matrix, 
i.e. $R=1$ in Eq.~(\ref{eq:seesaw:casas}). 
Leading to the results displayed in this section, we
have  taken into account  
all available LEP and Tevatron
bounds on the Higgs boson and SUSY spectrum~\cite{PDG,Higgs:LEP,Higgs:LHC}, 
as well as the most recent results on negative SUSY searches from 
the LHC collaborations~\cite{SUSY:ATLAS,SUSY:CMS}.

\noindent Concerning the WMAP7 bound for the observed dark matter relic 
density~\cite{WMAP},
\begin{equation}\label{exp:dm:wmap}
0.0941\, \lesssim \,\Omega h^2 \,\lesssim \, 0.1277\,,
\end{equation}
 we do not systematically impose it as a viability
requirement in our analysis. Nevertheless, we do 
require the lightest neutralino to be the lightest SUSY 
particle (LSP). We will return to this issue at a later stage.

\bigskip
Let us then begin our discussion by investigating how the requirements of
a ``standard window'', as well as compatibility with experimental
bounds, constrain the type III SUSY seesaw parameter space in the 
case of degenerate fermion triplets (i.e., $M_{N_i} = M_{24}$).
\begin{figure}[t!]
\begin{center}
\hspace*{-10mm}
\begin{tabular}{cc}
\epsfig{file=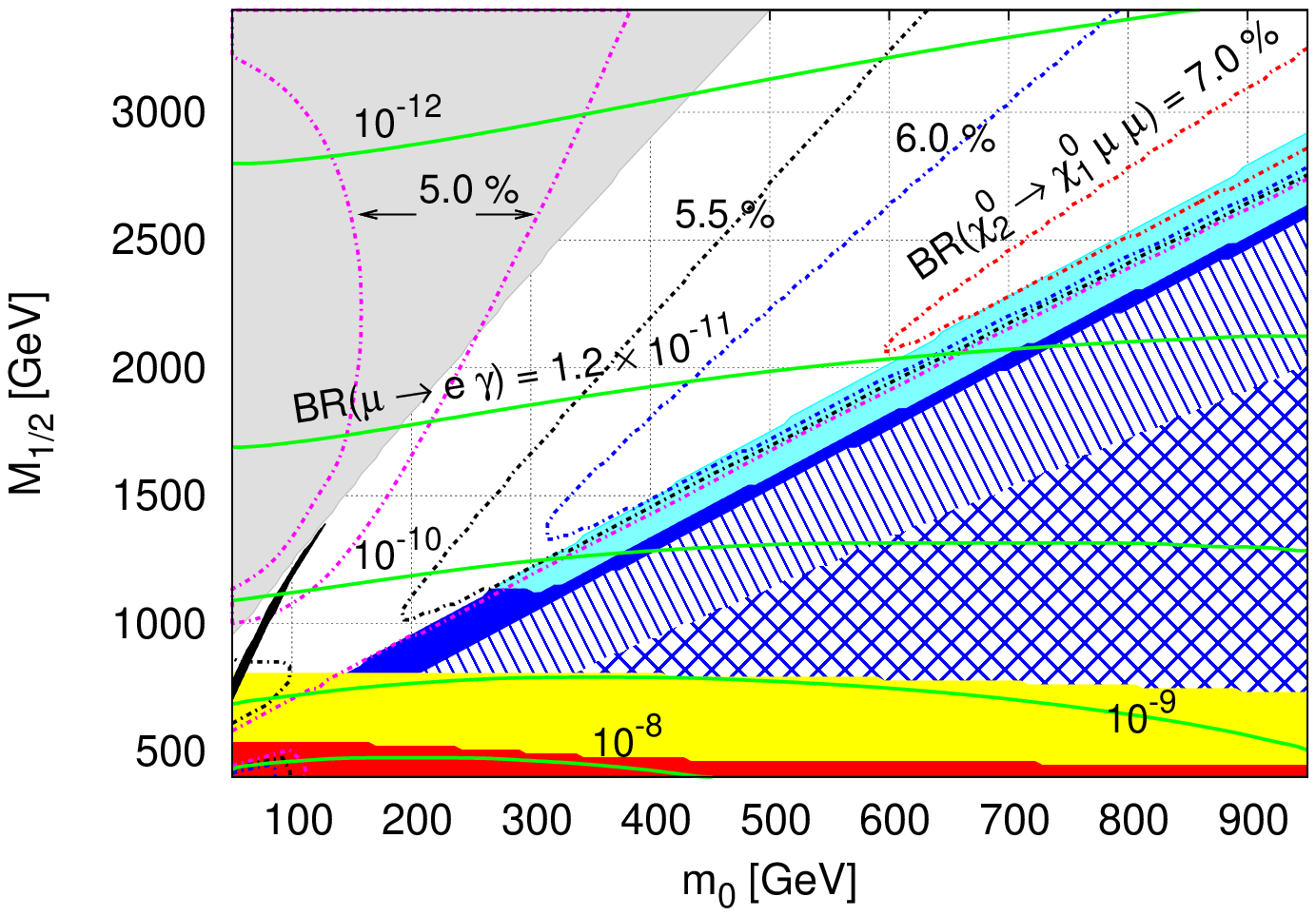, clip=, angle=0, width=85mm}&
\epsfig{file=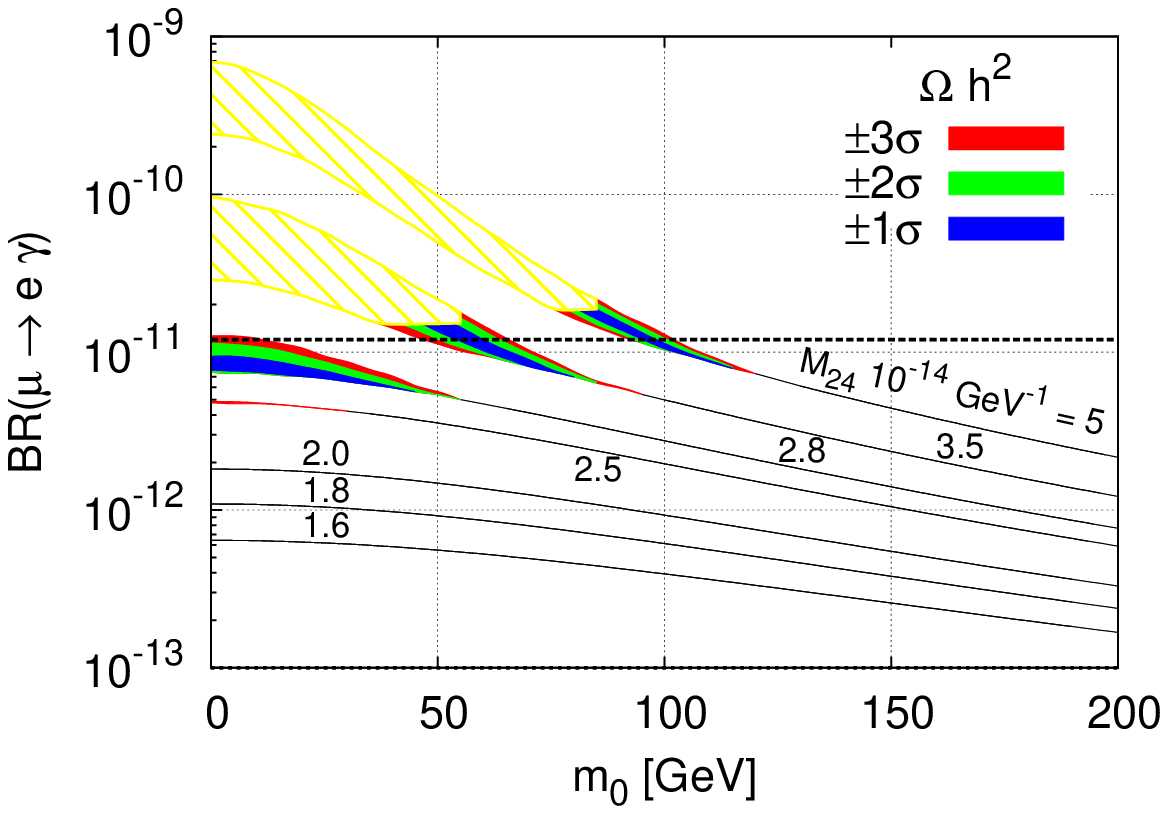, clip=, angle=0, width=85mm}
\end{tabular}
\caption{
On the left, $m_0 - M_{1/2}$ plane (in GeV), with $A_0=0$, $\tan
  \beta=10$, for a seesaw scale $M_{24} \sim 5 \times 10^{14}$ GeV and
  $\theta_{13}=0.1^\circ$. The shaded region on the left is excluded
  due to the presence of a charged LSP, while the yellow (red) region
  is excluded in view of $m_{h^0_1}$ bounds ($m_{h^0_1}$ and LHC bounds). 
  Several regions do not fulfil the ``standard window'' requirements: solid
  regions correspond to having $m_{\chi_2^0} < m_{\tilde \ell_{_L}}+ 10$
  GeV (cyan) and $m_{\chi_2^0} < m_{\tilde \tau_{_2}}+ 10$
  GeV (blue). The dashed blue region corresponds to $m_{\chi_2^0} <
  m_{\tilde \ell_{_L},\tilde \tau_{_2}}$ while blue crosses correspond
  to $m_{\chi_2^0} < m_{\tilde \tau_{_1}}+ m_{\tau}$.
  The centre white region denotes the parameter space obeying the
  ``standard window'' constraint. Green lines denote isocurves for 
  BR($\mu \to e\gamma$), while the dashed-dotted lines correspond to
  different values of BR$(\chi_2^0\to \chi_1^0\  \mu \mu)$
  as indicated in the plot.
  A small black region in the lower left corner
  corresponds to a WMAP7 compatible $\chi_1^0$ relic density. 
On the right panel, BR($\mu \to e \gamma$) as a function of $m_0$ (in GeV),
for $A_0=0$, $\tan \beta=5$, $\theta_{13}=0.1^\circ$ 
and several values of $M_{24}$: $1.6 \times 10^{14}$
GeV $\lesssim M_{24} \lesssim 5 \times 10^{14}$ GeV (from lower to
upper curves). Horizontal lines correspond to the current bound and future
sensitivity. The yellow gridded region is excluded due to violation of 
$m_{h^0_1}$ bounds. 
The colour code
denotes compatibility with the WMAP7 bounds on $\Omega h^2$.}
\label{fig:stdwindow}
\end{center}
\end{figure}

 On the left-hand side of Fig.~\ref{fig:stdwindow}, we display the 
$m_0- M_{1/2}$ parameter space for a type III SUSY seesaw, taking 
$A_0=0$, $\tan  \beta=10$, and a seesaw scale $M_{24} 
\sim 5 \times 10^{14}$ GeV, setting also $\theta_{13}=0.1^\circ$.
The excluded (shaded) areas correspond to a
charged LSP, to the violation of collider constraints on the Higgs
 and sparticle spectrum,  and to kinematically 
disfavoured regimes (kinematically
closed $\chi_2^0 \to \tilde \ell \ell$ channel, excessively soft
outgoing leptons, etc.). The requirements
of a ``standard window'' (see section~\ref{sec:lfv}) 
are fulfilled on the central white region. 
For this choice of SUSY seesaw parameters, 
a large part of the latter viable region
is excluded since it is associated with an excessively large $\mu \to e
\gamma$ branching ratio, as can be verified from the isocurves for
the BR($\mu \to e \gamma$). 
Additional isocurves (dashed-dotted 
lines) denote BR($\chi_2^0\to \chi_1^0 \ \mu \mu$). 
In the region complying with the ``standard window'' requirements, the
latter range from $5\%$ to $7\%$; for the LHC operating at
$\sqrt{s}=7$ TeV, hardly any events would be observable, 
while for $\sqrt{s}=14$ TeV, one could expect some 
$10 \text{ to } 1000$ events (for an integrated luminosity of 
$100 \text{ fb}^{-1}$). This implies that these $\chi_2^0$
decay chains could indeed be studied at the higher luminosity and higher
energy phase of LHC. 

Concerning dark matter, 
it is important to notice that, although the requirements imposed on the
$\chi_2^0 \to \tilde \ell \ell$ decay usually lead to a region where 
the correct dark matter relic density could in principle be obtained
from co-annihilations of the LSP with the next-to-LSP (NLSP), 
finding points for which $\Omega h^2$ is
indeed in agreement with WMAP7 data proves to be challenging.
For the particular SUSY seesaw
configuration investigated in Fig.~\ref{fig:stdwindow},
we verify that the regions where one finds 
the correct dark matter relic density
are already excluded due to having 
an excessively large BR($\mu \to e \gamma$).  
Although viable DM
scenarios in the type III SUSY seesaw are indeed very 
constrained~\cite{Esteves:2010ff},
regions can be found where either by a different choice of seesaw
parameters (e.g. setting $\delta$, the Dirac phase in $U_\text{MNS}$, 
$\delta=\pi$) or for 
smaller $\tan \beta$ values, a viable $\Omega h^2$ can be obtained, but 
still associated with a considerable fine tuning of the parameters.
This is illustrated on the right-hand side plot of
Fig.~\ref{fig:stdwindow} for $\tan\beta = 5$,
where we display BR($\mu \to e \gamma$) as a function of $m_0$ for
several (7) choices of  the seesaw
scale, $1.6 \times 10^{14}$ GeV $\lesssim M_{24} \lesssim 5 \times
10^{14}$ GeV.  When compatibility with 
the  WMAP7 $3\sigma$ interval for $\Omega h^2$ is indeed possible, 
$M_{1/2}$ has been varied (corresponding to the coloured solid regions
as well as the gridded ones - which are already excluded 
by collider constraints); else, we display the value 
of  $M_{1/2}$ that minimises the deviation from the WMAP7 $3\sigma$
interval (black curves).  
Typically, the correct relic density 
is obtained for nearly degenerate LSP and NLSP.

\begin{figure}[t!]
\begin{center}
\hspace*{-10mm}
\begin{tabular}{cc}
\epsfig{file=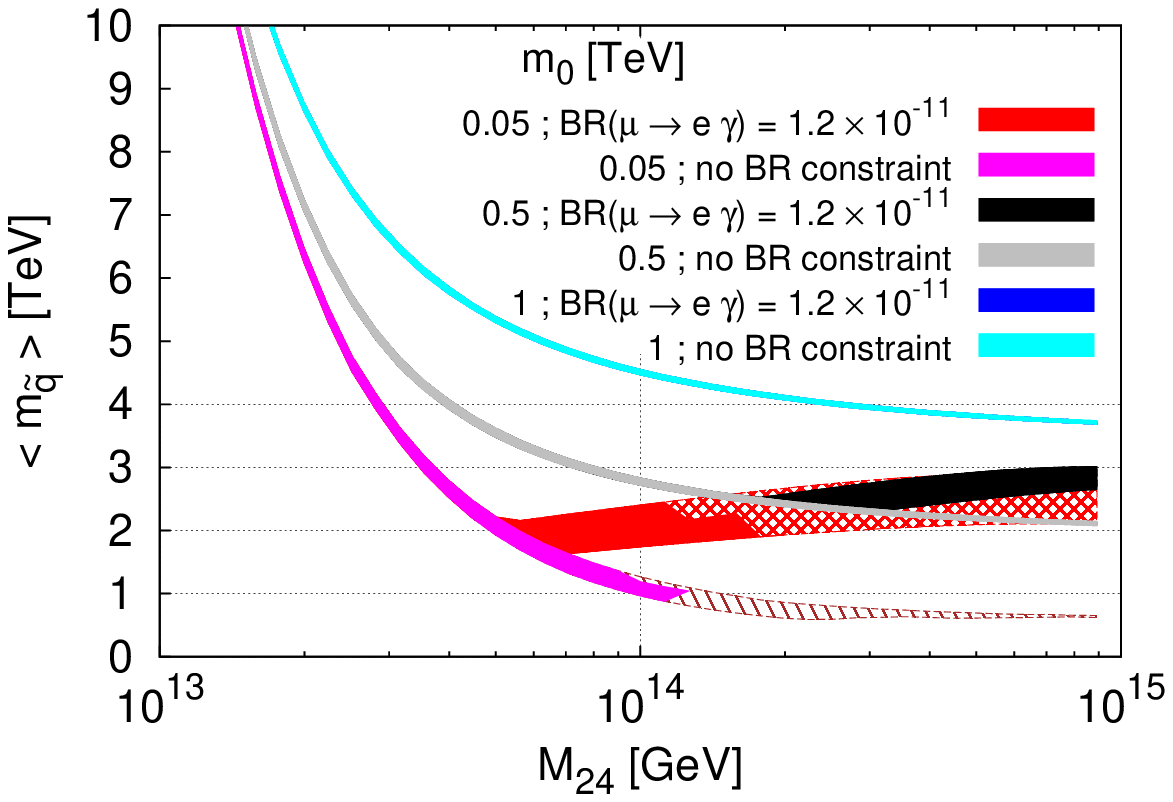, clip=, angle=0, width=85mm}&
\epsfig{file=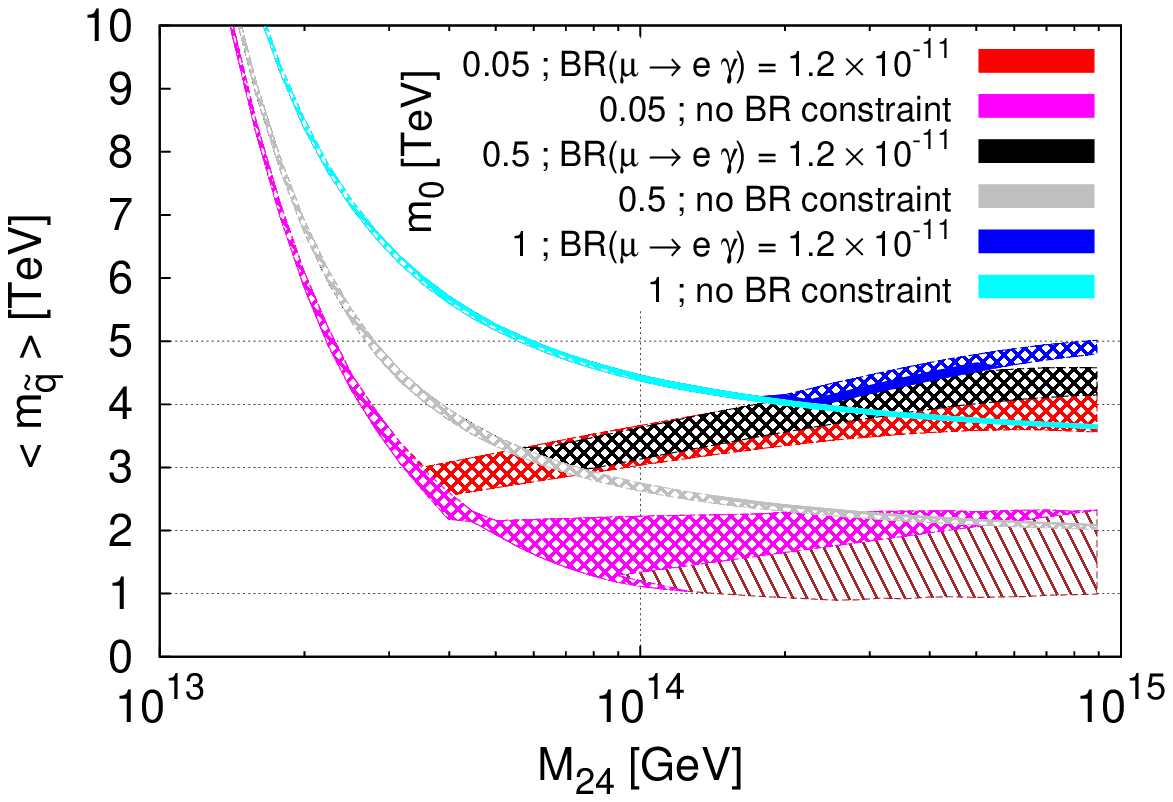, clip=, angle=0, width=85mm}
\end{tabular}
\caption{Average squark mass range (in TeV) as a function of the
  triplet mass (in GeV), for different values of $m_0$: 50 GeV (blue/cyan),
500 GeV (black/grey) and 1 TeV (red/pink), the colour code
further denoting imposing/not imposing the bound on BR($\mu \to e
\gamma$). Gridded regions correspond to cases where one has a charged
LSP. The brown region is excluded due to violation of LHC or
$m_{h^0_1}$ bounds. 
The left (right) figure corresponds to $\tan \beta=10\,(40)$. In both cases,
$\theta_{13}=0.1^\circ$, $A_0=\{-1,0,1\}$ TeV, 
with $M_{1/2}$  set to the lowest possible
value complying with the requirement of a ``standard window''.}
\label{fig:avsquark:M24}
\end{center}
\end{figure}

Contrary to the type I seesaw, where the requirements of observing the 
$\chi_2^0 \to \chi_1^0\, \ell\, \ell  $ chain did not significantly
alter the expected low-energy SUSY spectrum,
important changes are expected in the type III seesaw, especially due to the (strong)
running of the gaugino masses. Moreover, and as discussed previously, the 
allowed interval for the triplet masses ($M_{24}$) is also severely 
constrained. To illustrate the impact of a ``standard window'' on the
spectrum, we display 
in Fig.~\ref{fig:avsquark:M24} the (geometrically)
averaged squark masses as a function of the triplet mass, for
different values of $m_0$. We consider two regimes of $\tan
\beta$, $\tan
\beta=10$, 40. For each point a scan over $M_{1/2}$ is
conducted to determine its lowest possible
value complying with the requirement of a ``standard window''.
We also differentiate between the ranges allowed with and without applying the
current bound on BR($\mu \to e \gamma$). Regarding mixings in the 
neutrino sector, we again work in the
limit $R=1$ and set $\theta_{13}=0.1^\circ$.

As can be seen from Fig.~\ref{fig:avsquark:M24}, and 
as hinted on section~\ref{sec:SUSYtypeIII}, the allowed
interval for the seesaw scale (here represented by $M_{24}$)
ranges from $10^{13}$ GeV to just below
$10^{15}$ GeV, corresponding to the results of~\cite{Esteves:2010ff}.
It is worth emphasising that there are regions where,
in addition to complying with all accelerator and neutrino data, the
type III seesaw still leads to scenarios of LFV in agreement with
low-energy data (the most stringent constraint arising from the $\mu \to e
\gamma$ decay). This diverges from the findings
of~\cite{Esteves:2010ff}, where only very light SUSY spectra were considered.  
Regimes of heavier sparticles (large $M_{1/2}$ and $m_0$) are
preferred, confirming that these scenarios would be more likely to be observed 
at the LHC for $\sqrt{s}=14$ TeV.
It is important to remark that, even for a regime of small $m_0$, 
we are always led
to a very heavy SUSY spectrum (here
represented by a geometrical average of the squark masses). Complying
with all the above requirements implies that even for $m_0$ as low as 50
GeV, one must have $<m_{\tilde q}>^\text{min} \sim 2$ TeV (and around 1.5 TeV for
the limiting case of
$m_0=0$). By itself, this result is important in the sense that should
any light SUSY spectrum be discovered at the LHC in association with
the $\chi_2^0 \to \tilde \ell \ell$ decay chain, this would strongly
suggest that a type III seesaw is not at work. 
It is also important to notice that the steep increase of $<m_{\tilde
  q}>$ for lower values of $M_{24}$ is a direct consequence of 
having imposed the requirement of a ``standard window''. In particular
the strong running of $M_2$ would imply that for lower $M_{24}$ the
mass of the sleptons would be much larger than that of the
neutralinos, thus preventing the cascade decay $\chi_2^0 \to \tilde
\ell \ell$.

\begin{figure}[h!]
\begin{center}
\hspace*{-10mm}
\begin{tabular}{cc}
\epsfig{file=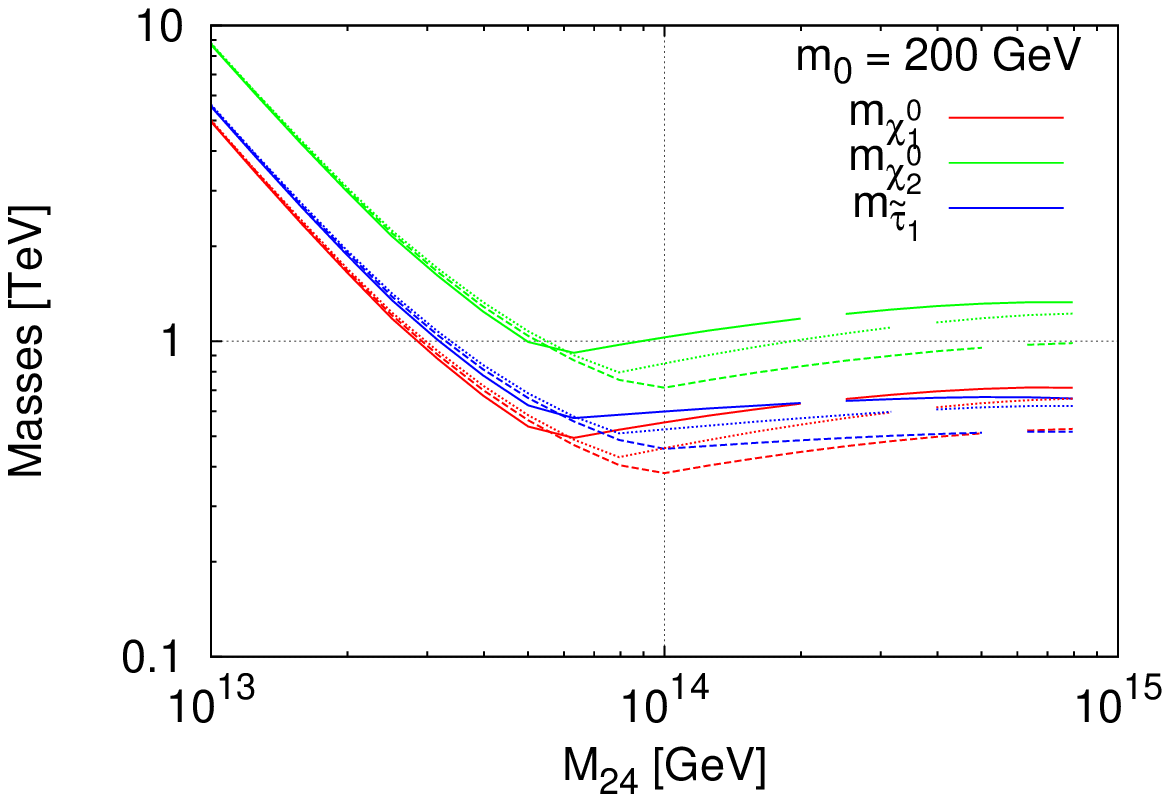, clip=, angle=0, width=85mm}&
\epsfig{file=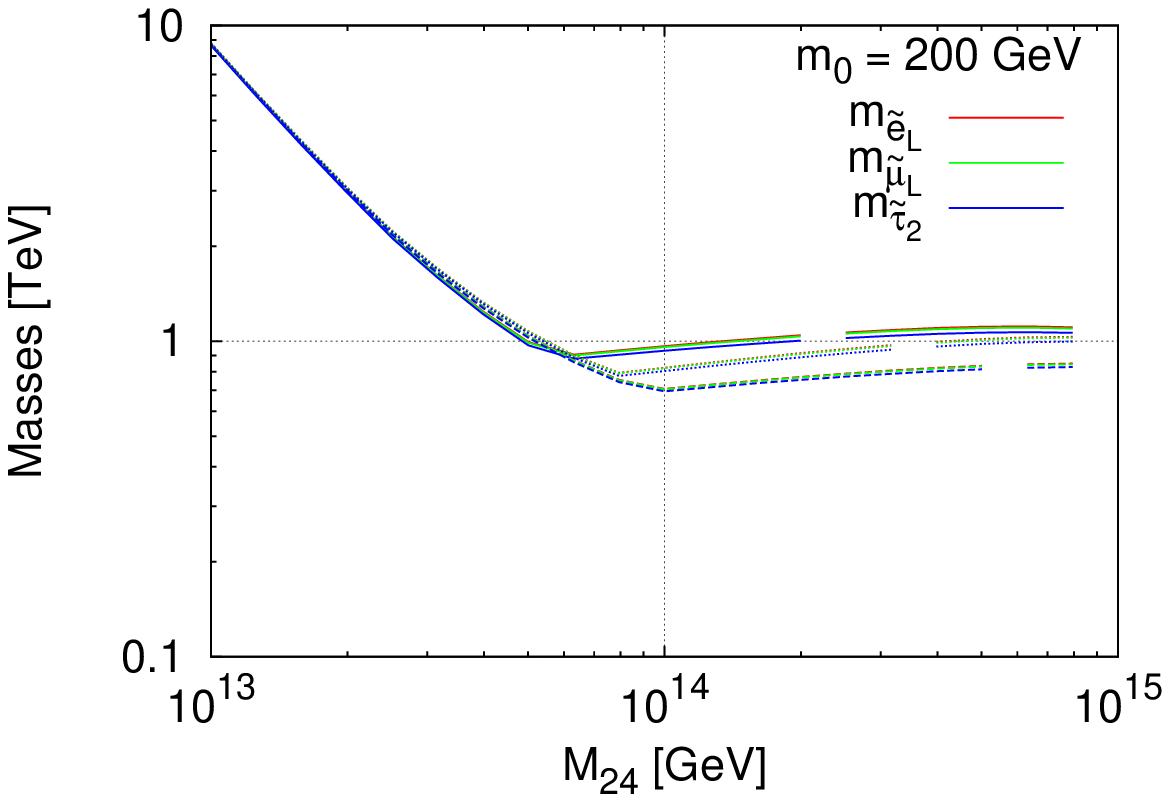, clip=, angle=0, width=85mm}
\end{tabular}
\caption{Gaugino and slepton masses (in TeV) as a 
function of the triplet mass, $M_{24}$
  (in GeV), for $m_0=$ 200 GeV. On the left, $m_{\chi_{1,2}^0}$ and 
  $m_{\tilde \tau_{_1}}$; on the right $m_{\tilde e_{_L}}$, 
  $m_{\tilde \mu_{_L}}$ and $m_{\tilde \tau_{_2}}$. We have taken  
  $\tan \beta=10$ and set $\theta_{13}=0.1^\circ$.  
  $M_{1/2}$ is set to the lowest possible
  value complying with the requirement of a ``standard window'' and
  with the bound on BR($\mu \to e \gamma$). In both cases, the
  different lines correspond to distinct values of $A_0$: -1 TeV (full), 
  0 (dashed), and 1 TeV (dotted). An interrupted line signals the
  onset of a charged LSP region. }
\label{fig:mass:emu:M24}
\end{center}
\end{figure}

Increasing the value of $\tan \beta$ has an effect on the SUSY
contributions to the LFV observables (which grow with $\tan^2 \beta$,
see Eq.~(\ref{eq:BR:MIA:LL})), implying that larger values of the SUSY
spectrum (and hence of $M_{1/2}$) are required to comply with the
experimental constraints. Furthermore, the augmentation of the $LR$
mixing in the stau sector implies that having a neutral LSP becomes
increasingly difficult. For $\tan \beta=40$, as depicted on the 
right-hand side
of Fig.~\ref{fig:avsquark:M24}, the allowed regions are extremely reduced:
only a thin blue band (corresponding to $m_0=1$ TeV) survives all constraints.
To further clarify and illustrate the above discussion regarding the
dependence of the sparticle spectrum on the seesaw
scale (under the requirements of a ``standard window'' and
compatibility with experimental bounds), we present on 
Fig.~\ref{fig:mass:emu:M24} the electroweak gaugino and slepton masses as a 
function of the triplet mass ($M_{24}$), also explicitly denoting the
value of $A_0$ in each case. Being essentially driven by $M_{1/2}$,
the running of their values is similar to that of the (averaged) squark
masses. 

Finally, let us notice that 
variations of the still unknown Chooz angle, $\theta_{13}$, have a
comparatively small impact: they only contribute to some of the LFV
observables and compatibility with the experimental bound is easily
recovered through a minor augmentation of $M_{1/2}$, which in turn 
leads to a heavier sparticle spectrum (for fixed values of $m_0$).

\begin{figure}[t!]
\begin{center}
\hspace*{-10mm}
\begin{tabular}{cc}
\epsfig{file=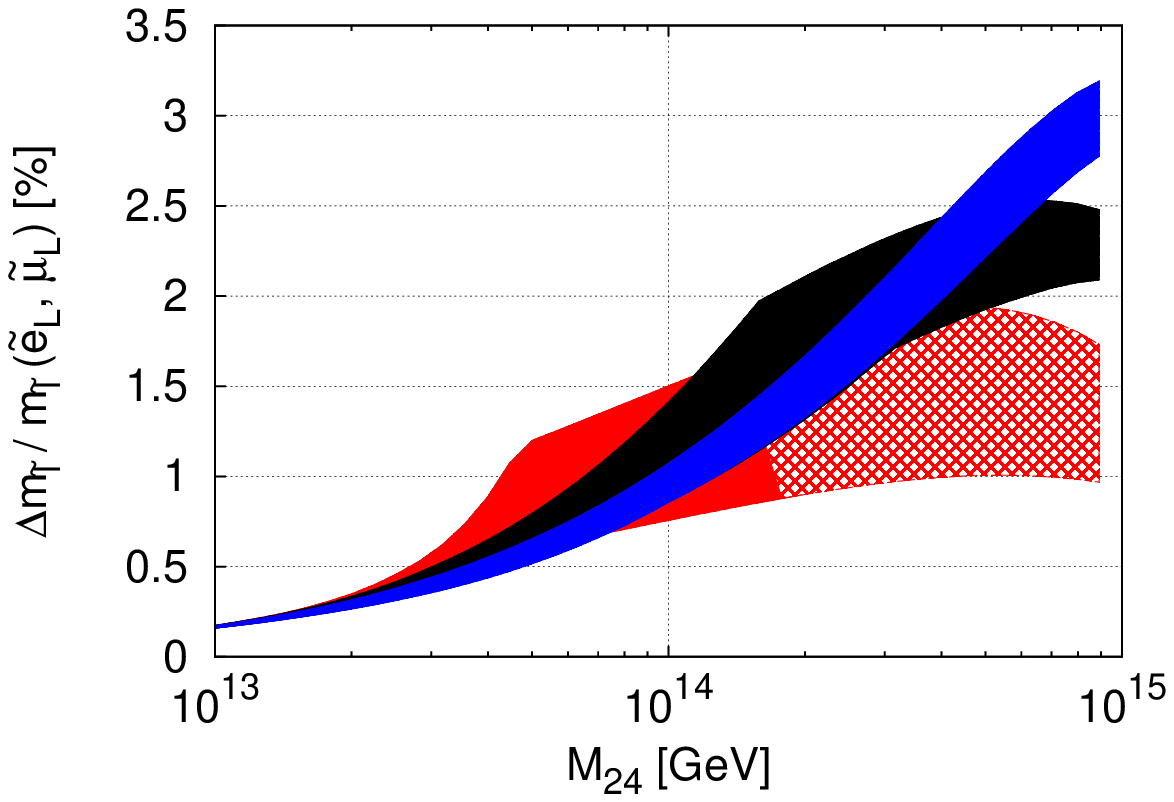, clip=, angle=0, width=85mm}&
\epsfig{file=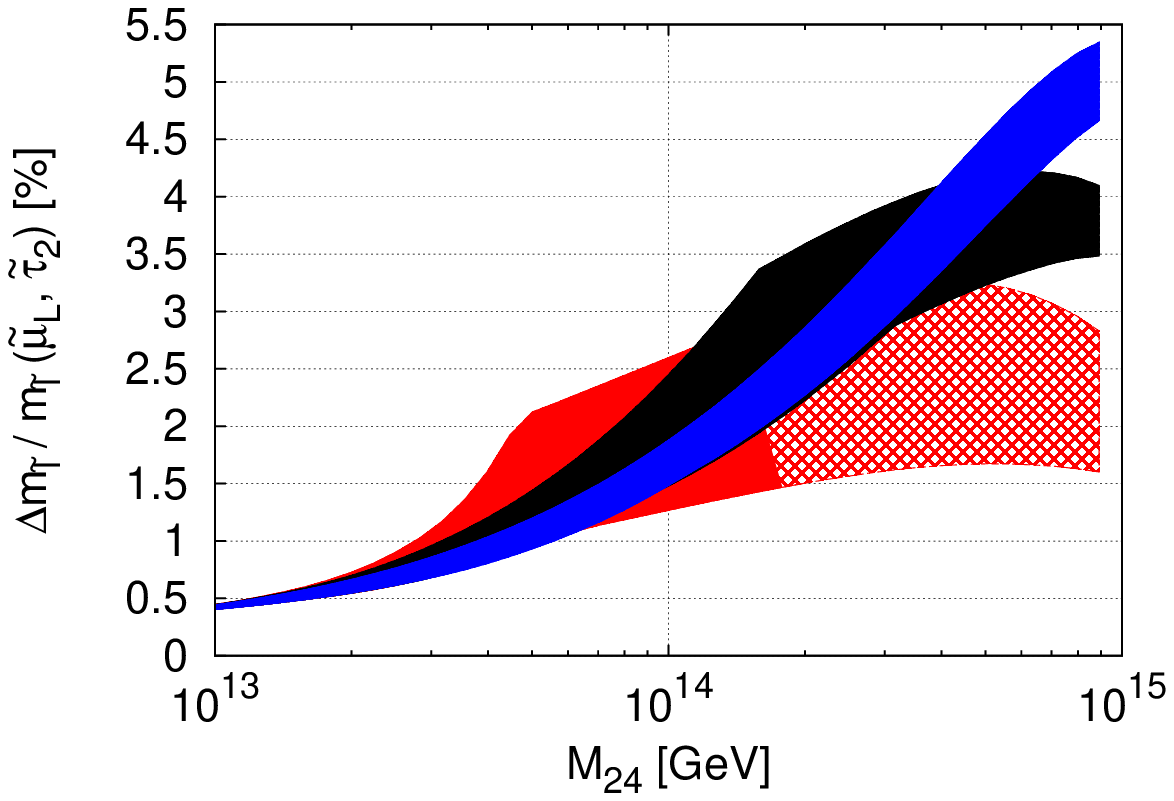, clip=, angle=0, width=85mm}
\end{tabular}
\caption{On the left, $\tilde e_{_L} - \tilde \mu_{_L}$ 
  mass difference  
  (normalised to an average slepton mass) 
  as a function of the triplet mass, $M_{24}$ (in GeV). 
  On the right, $\tilde  \mu_{_L} - \tilde \tau_{_2}$ effective mass
  difference (normalised to the
  corresponding average slepton mass) also as a function of the
seesaw scale.  In both cases we take
  $\tan \beta=10$, $\theta_{13}=0.1^\circ$, and consider 
  different values of $m_0$: 50 GeV (red), 500 GeV (black),
  and 1 TeV (blue). Gridded regions correspond to a charged LSP. 
For each point one varies $A_0=\{-1, 0, 1\}$ TeV, 
  while $M_{1/2}$ is set to the lowest possible
  value complying with the requirement of a ``standard window'' and
  with the bound on BR($\mu \to e \gamma$).}
  \label{fig:deltaMS:emu:M24}
\end{center}
\end{figure}

\bigskip
We now focus our discussion on the slepton mass differences, as
potentially measurable at the LHC. We recall that the expected
(conservative) sensitivities for the slepton mass splittings are 
of $\mathcal{O}(0.1\%)$ for $\Delta m_{\tilde
\ell}/m_{\tilde \ell}\  (\tilde e ,\tilde \mu)$ and $\mathcal{O}(1\%)$
for $\Delta m_{\tilde \ell}/m_{\tilde \ell}\  (\tilde \mu , \tilde
\tau)$.
In Figs.~\ref{fig:deltaMS:emu:M24} we display the slepton mass
splittings (effective mass difference in the case of 
$\tilde  e_{_L} -\tilde \mu_{_L}$) as a function of the seesaw scale, for the same
parameter scan as in Figs.~\ref{fig:avsquark:M24}. One 
verifies that 
$\Delta  m_{\tilde{\ell}}/m_{\tilde{\ell}} \ ({\tilde e_{_L}}, {\tilde \mu_{_L}})$ 
can be as large as 3\%
and $\Delta  m_{\tilde{\ell}}/m_{\tilde{\ell}} \ 
({\tilde \mu_{_L}}, {\tilde \tau_{_2}}) \sim 5\%$, 
for the maximal values
of the seesaw scale, and for large $m_0$ regimes (where the largest
amount of flavour violation, still compatible with experimental
bounds and with the requirements of a ``standard
window'', occurs).
For larger values of $\tan \beta$, one could have slightly larger 
$\tilde  \mu_{_L} - \tilde \tau_{_2}$ mass splittings (mostly in association 
with larger $LR$ mixings in the stau sector), but the viable regions
in the parameter space are much smaller, as mentioned before. In all 
cases one always has 
$\Delta  m_{\tilde{\ell}}/m_{\tilde{\ell}} 
\ ({\tilde \mu_{_L}}, {\tilde \tau_{_2}}) \lesssim 7\%$. 

When compared to a type I SUSY seesaw (see~\cite{Abada:2010kj}), one
realises that the maximal values of the slepton mass splittings are
slightly smaller, which is a consequence of the somewhat heavier SUSY
spectrum. Concerning the mass splittings of
right-handed sleptons, and analogous to the type I case, 
one finds  a very small effect:
in fact, for the parameter space surveyed in
Fig.~\ref{fig:deltaMS:emu:M24} (and always under the imposition of a
``standard window" as well as compatibility with collider constraints),
 $\Delta  m_{\tilde{\ell}}/m_{\tilde{\ell}} \ ({\tilde \mu_{_R}}, 
{\tilde e_{_R}}) \lesssim 0.1\%$.

Finally, assuming that slepton mass differences are measured close to
their maximal values ($\sim 3\%$ for $\tilde \mu_{_L} - \tilde e_{_L}$ and 
$\sim 5\%$ for $\tilde \mu_{_L} - \tilde \tau_{_2}$) and that the reconstructed
value of $m_0$ is found to be large (around 1 TeV) then, as seen from
Figs.~\ref{fig:deltaMS:emu:M24},
this would suggest that the seesaw scale would be $M_{24}\sim
10^{15}$ GeV (for the limiting case $R=1$).

\begin{figure}[h!]
\begin{center}
\hspace*{-10mm}
\begin{tabular}{cc}
\epsfig{file=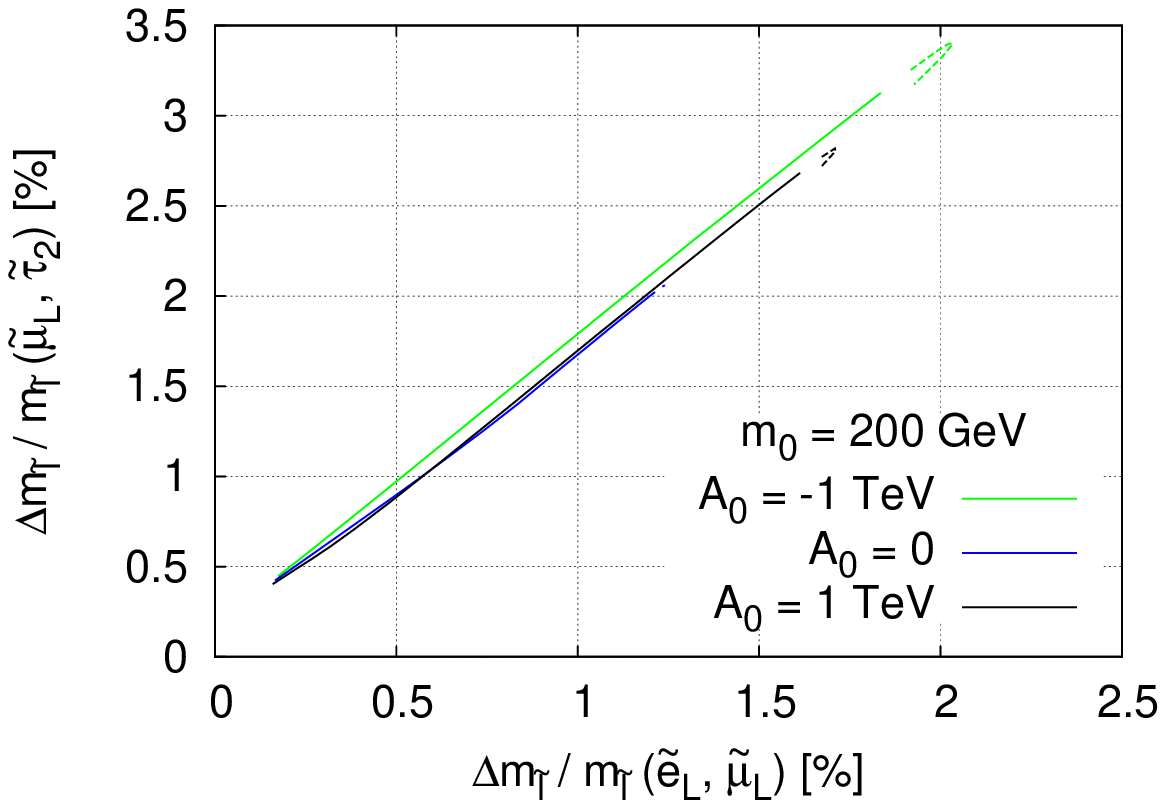, clip=, angle=0, width=85mm} &
\epsfig{file=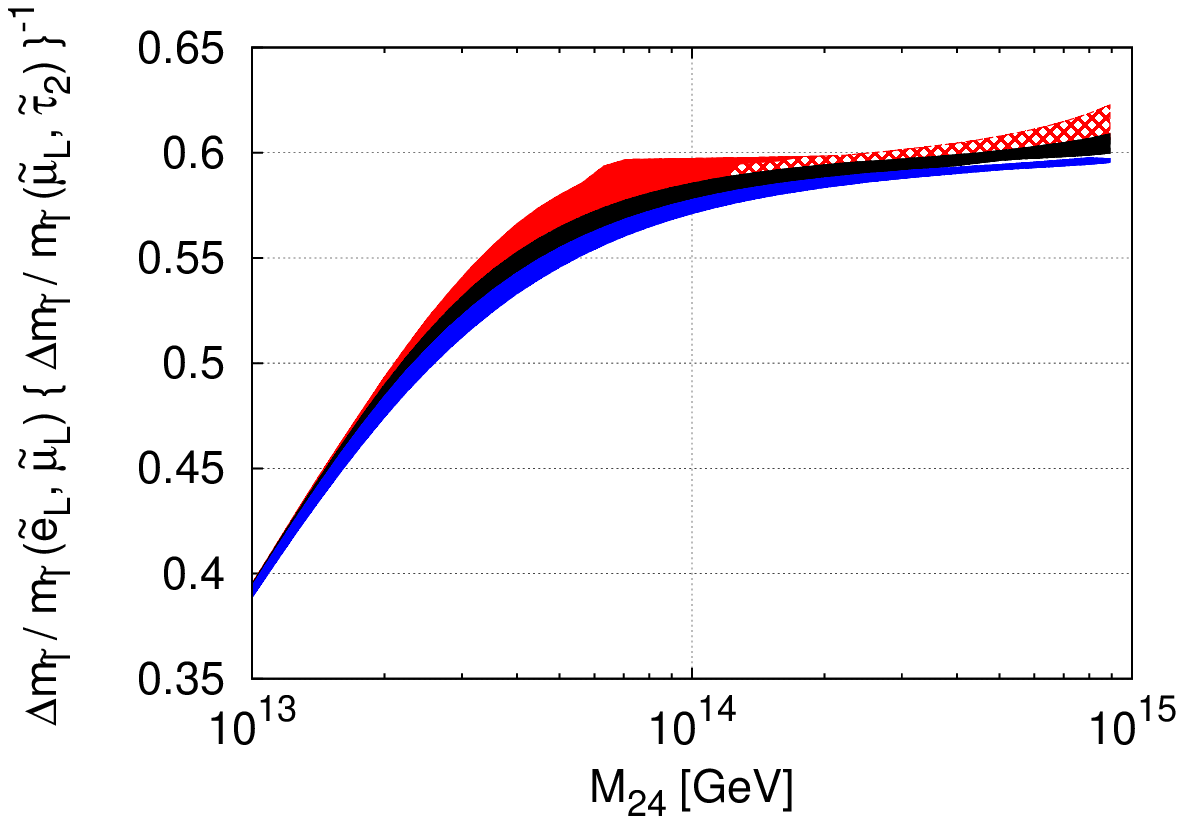, clip=, angle=0, width=85mm} 
\end{tabular}
\caption{On the left, 
$\Delta m_{\tilde \ell}/m_{\tilde \ell}\ (\tilde  \mu_{_L} , \tilde \tau_{_2})$
as a function of $\Delta m_{\tilde \ell}/m_{\tilde \ell}\ (\tilde e_{_L} ,
\tilde \mu_{_L})$, for $m_0=200$ GeV, $\tan \beta=10$, $\theta_{13}=0.1^\circ$
and taking $A_0=\{-1,0,1\}$ TeV (green, blue and black lines,
respectively). An interrupted (dashed) line signals the onset of 
a charged LSP regime towards larger values of the mass splittings.
On the right, ratio of slepton mass differences, $\Delta m_{\tilde
    \ell}\ (\tilde e_{_L} , \tilde \mu_{_L}) / \Delta m_{\tilde
    \ell}\ (\tilde  \mu_{_L} , \tilde \tau_{_2})$ (normalised to the
 corresponding average slepton mass), as a function of the
triplet mass (in GeV), for different values of $m_0$, with 
$\tan \beta=10$, $A_0=\{-1, 0, 1\}$ TeV
and $\theta_{13}=0.1^\circ$. Scan and colour code as in
Fig.~\ref{fig:deltaMS:emu:M24}.}
\label{figdeltaMS:emu.mutau:M24}
\end{center}
\end{figure}

In Fig.~\ref{figdeltaMS:emu.mutau:M24} we present the comparison of
the $\tilde e_{_L} - \tilde \mu_{_L}$ and $\tilde \mu_{_L} - \tilde \tau_{_2}$
mass differences, as well as their ratio, as a function of the seesaw
scale. Similar to what occurs for a type I SUSY seesaw, and as
discussed in Section~\ref{sec:lfv}, 
the mass differences are strongly correlated (being
driven by the $(\Delta m^2_{\tilde L})_{_{23}}$ entry in the slepton mass
matrix).
With the exception of the regions corresponding to smaller values of
$M_{24}$, 
the relation $\Delta m_{\tilde \ell}/m_{\tilde \ell}\ (\tilde e_{_L} ,
\tilde \mu_{_L}) \approx \mathcal{O}(1/2) 
\Delta m_{\tilde \ell}/m_{\tilde \ell}\ (\tilde \mu_{_L} , \tilde \tau_{_2})$
(Eq.~(\ref{eq:MS:mutau:emu:R1seesaw}))
typically holds to a very good approximation
(with corrections due to fact that flavour conserving radiative
corrections driven by the tau Yukawa coupling now play a non-negligible
r\^ole). For lower values of the seesaw scale, 
where the requirement of a ``standard window''
(i.e. $\chi_2^0 \to \tilde \ell \ell$ decay, with hard outgoing
leptons) forces a rapid increase of $M_{1/2}$, 
a small deviation to this strict
correlation is observed. This can also be seen in the left-hand side
of Fig.~\ref{figdeltaMS:emu.mutau:M24}, zooming into the lower
end of the lines. 
We have verified that this behaviour occurs irrespective of
the value of $\theta_{13}$ and for all $\tan \beta$ regimes (provided
that the regions are phenomenologically and experimentally viable).

\begin{figure}[t!]
\begin{center}
\hspace*{-10mm}
\begin{tabular}{cc}
\epsfig{file=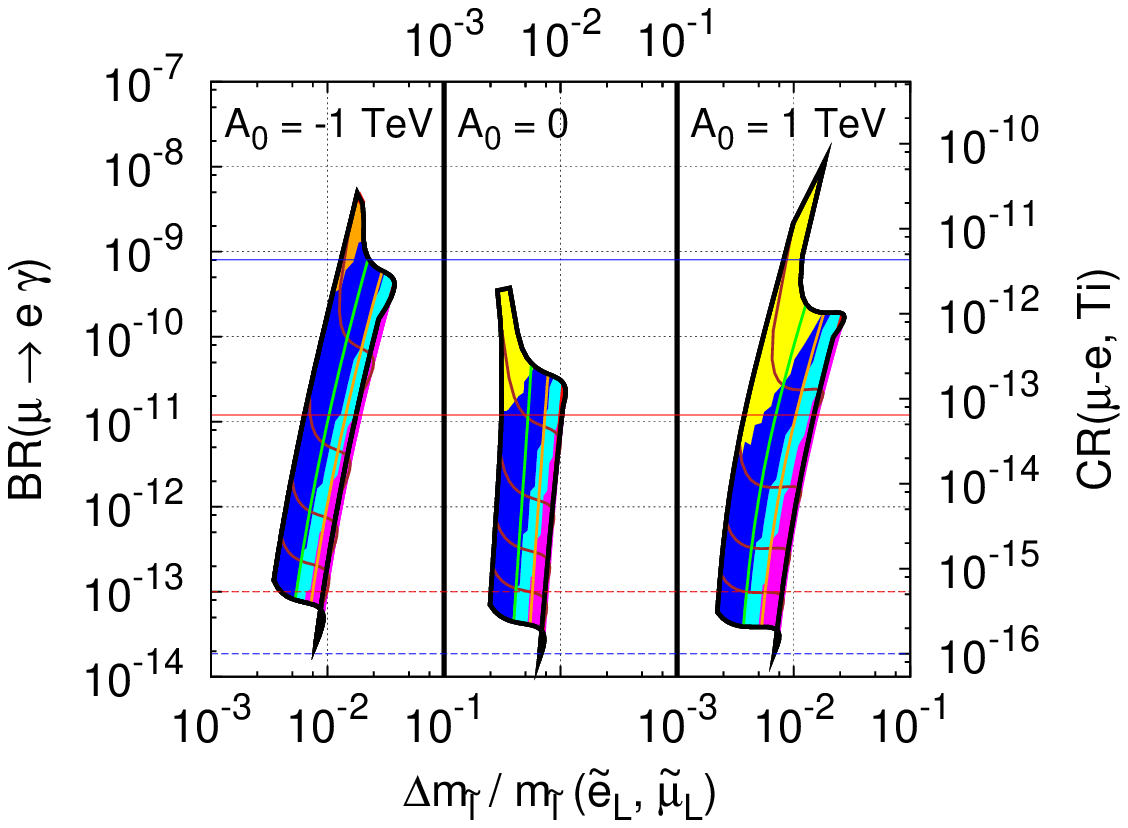, clip=, angle=0, width=88mm}&\hspace*{-10mm}
\epsfig{file=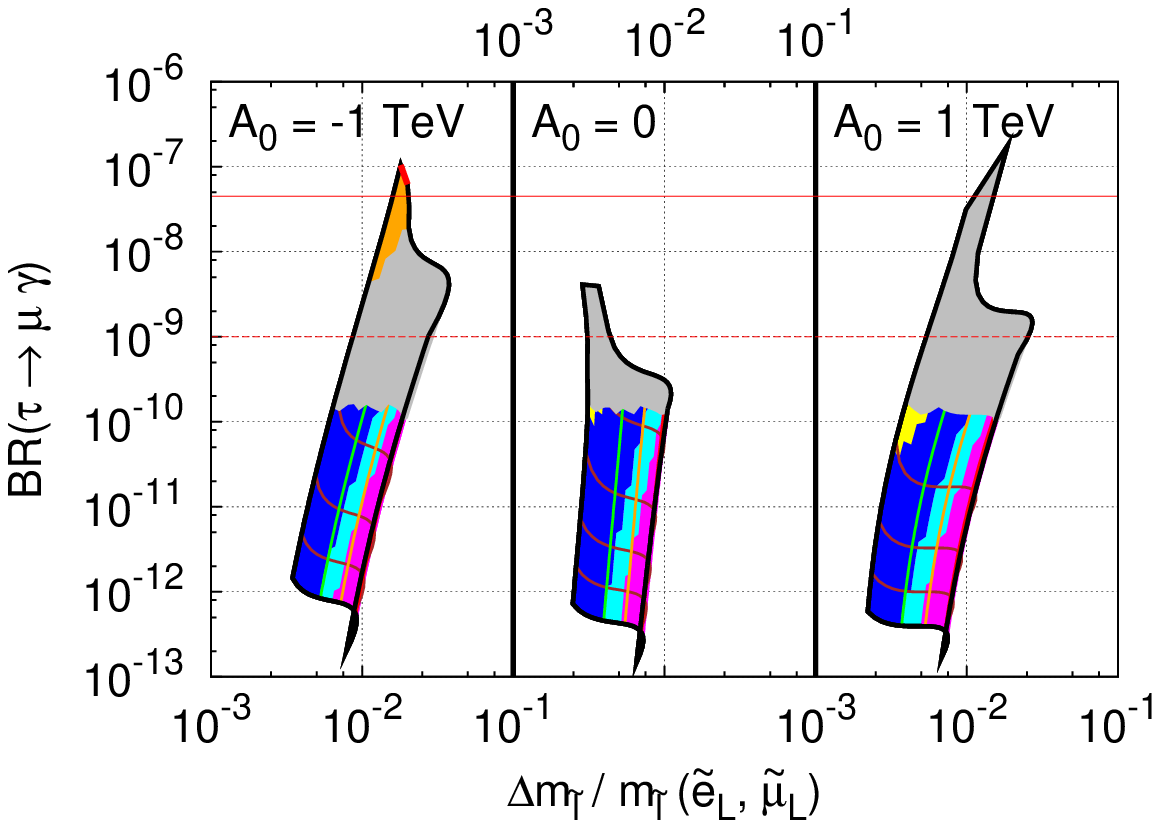, clip=, angle=0, width=88mm}
\end{tabular}
\caption{BR($\mu \to e \gamma$) and BR($\tau \to \mu \gamma$) 
  as a function of the $\tilde e_{_L} -
  \tilde \mu_{_L}$ slepton mass difference (normalised to an
  average slepton mass), corresponding to the left- and right-hand side
  panels. In both cases we have set  $m_0=100$ GeV, $\tan \beta=10$ and
  $\theta_{13}=0.1^\circ$, and considered different values of the triplet scale
  $M_{24}$ and of $M_{1/2}$. Each sub-panel corresponds to a distinct
  choice of $A_0$. Cyan regions correspond to fulfilling the
  requirements of a ``standard window''. 
  The bounds on $m_{h^0_1}$ are violated in the yellow regions,  
  LHC bounds on SUSY spectrum are violated in orange 
  regions,  while red regions are excluded due to both.
  Further excluded regions are due to failing to meet the
  kinematical constraints (blue), having a charged LSP (magenta) or
  violating another LFV bound (grey).
  Inset into each plot are ``horizontal'' isolines for $M_{1/2}$
  (ranging from 1.5 TeV to 6 TeV, from top to bottom) and ``vertical''
  isolines for $M_{24}$: from left to right,  
  $10^{13}$ GeV to  $9 \times 10^{14}$ GeV. 
  The secondary y-axis on the left-hand panel illustrates the
  corresponding values of CR($\mu - e$, Ti). Horizontal lines denote
  the current experimental bounds (full) and future sensitivities
  (dashed).}
\label{fig:BR:deltaMS:emu}
\end{center}
\end{figure}

The correlation of low- and high-energy LFV observables is explored 
in Fig.~\ref{fig:BR:deltaMS:emu}, where we present BR($\mu \to e \gamma$)
and BR($\tau \to \mu \gamma$)
as a function of the $\tilde e_L - 
  \tilde \mu_L$ slepton mass difference, taking $m_0=100$ GeV, and 
considering different values of the triplet scale,
$M_{24}$. We also provide additional information about the CR($\mu - e$,
Ti). As seen from both panels of Fig.~\ref{fig:BR:deltaMS:emu}, only a
small region of the scanned parameter space complies with the
requirements of a ``standard window'' while being in agreement with the several
experimental and phenomenological constraints. 
Similar to what occurs for a type I SUSY
seesaw, larger, negative values of $A_0$ translate into larger mass
splittings. 
The maximal amount of flavour violation, both regarding
radiative decays and slepton mass splittings, is obtained for: (i) a
seesaw scale as large as possible (without violating 
perturbativity arguments, specifically on $Y^\nu$), as
can be understood from Eqs.~(\ref{eq:seesaw:casas},
\ref{eq:LFV:LLog}); (ii) lower values of $M_{1/2}$ (leading to a
lighter SUSY spectrum, see Eq.~(\ref{eq:BR:MIA:LL})). Regarding the
$\tau \to \mu \gamma$ decays, as can be seen from the right panel of  
Fig.~\ref{fig:BR:deltaMS:emu}, the regions in parameter space associated 
with BR($\tau \to \mu \gamma$) 
within the sensitivity of SuperB are in fact excluded by the
present bounds on $\mu \to e \gamma$ decays.
Although we do not present the corresponding results, a similar study
with $m_0=1$ TeV leads to scenarios of somewhat larger mass splittings,
 and smaller
branching ratios for the radiative decays (due to the much heavier
spectrum).  It is nevertheless
interesting to remark that in this regime of very large $m_0$, one can have
maximal mixings in the lightest slepton - now a composition of
$\tilde \tau_{_L}$, $\tilde \tau_{_R}$ and $\tilde \mu_{_L}$ - possibly leading
to scenarios of very large mass splittings (albeit for a tiny
fraction of the parameter space).

\bigskip
Assuming that a type III seesaw is indeed the only source of LFV, and
given the extremely constrained parameter space, one finds that in
the conservative case of $R=1$, the corresponding slepton mass
splittings will always lie around the \% level, and are thus
within the expected sensitivity of the LHC. Furthermore, these mass
splittings correspond to values of BR($\mu \to e \gamma$) well within
the expected sensitivity of MEG (or even already ruled out by current
searches).
Moreover, the regions lying below MEG sensitivity have an associated 
CR($\mu - e$, Ti) 
within the reach of future experiments (PRISM/PRIME).

\bigskip
\begin{figure}[t!]
\begin{center}
\hspace*{-10mm}
\begin{tabular}{cc}
\epsfig{file=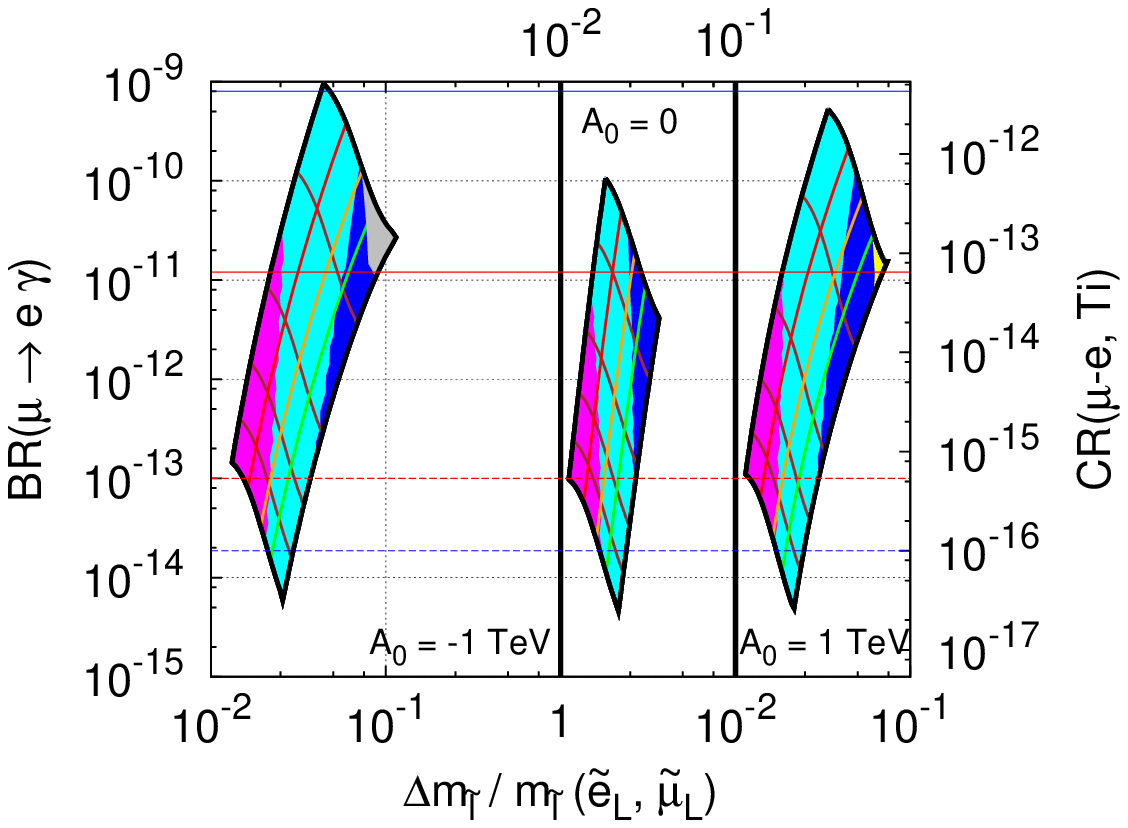, clip=, angle=0, width=88mm}&\hspace*{-10mm}
\epsfig{file=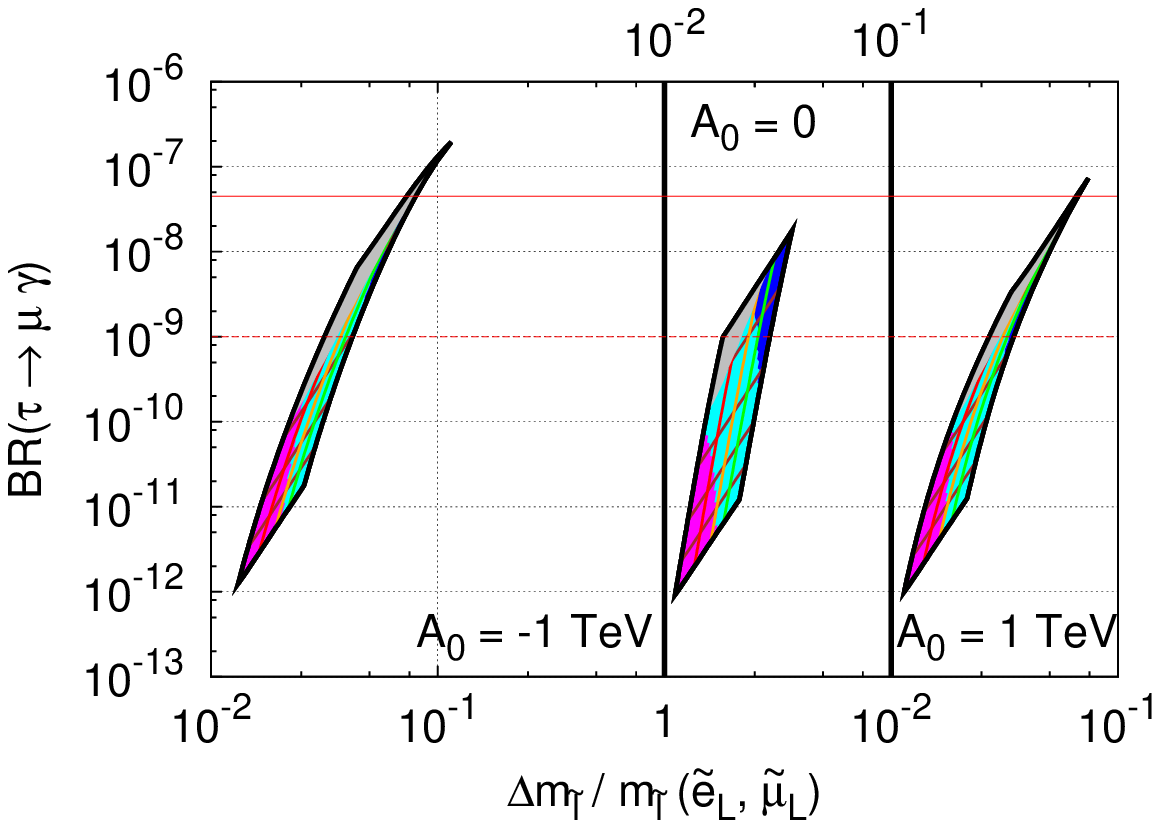, clip=, angle=0, width=88mm}
\end{tabular}
\caption{Non-degenerate triplet masses: BR($\mu \to e \gamma$) 
  and BR($\tau \to \mu \gamma$) 
  as a function of the $\tilde e_{_L} -
  \tilde \mu_{_L}$ slepton mass difference (normalised to an
  average slepton mass), corresponding to the left- and right-hand side
  panels. Same scan as leading to Fig.~\ref{fig:BR:deltaMS:emu}, except
  that now $M_{N_{1,3}}$ are fixed, with varying $M_{N_{2}}$: 
  $M_{N_{1}} = 10^{13}$ GeV$ \lesssim M_{N_{2}}\lesssim 9 \times 10^{14}$ 
  GeV $= M_{N_{3}}$.
  Same line and colour code as in
  Fig.~\ref{fig:BR:deltaMS:emu}, the only exception being that the
  inset ``vertical''
  isolines for $M_{N_{2}}$ {\it decrease} from left to right. }
\label{fig:NDmasses:BR:deltaMS:emu}
\end{center}
\end{figure}
We now consider more general scenarios of non-degenerate spectrum
for the heavy triplets. 
In order to investigate this regime, we fix the heaviest
(lightest) triplet mass to the upper (lower) limits of the $M_{24}$
interval previously obtained, and allow the next-to-lightest triplet
mass to vary between the latter limits. 
For such a non-degenerate triplet spectrum, we display in 
Fig.~\ref{fig:NDmasses:BR:deltaMS:emu} an analogous study to that 
of Fig.~\ref{fig:BR:deltaMS:emu} 
(same choice of the SUSY parameters, still working in 
the limiting case of $R=1$). 
As can be observed, the area complying with all
requirements (cyan band) is now comparatively larger. The $\tilde e_{_L} -
\tilde \mu_{_L}$ slepton mass differences are also 
enhanced when compared to the degenerate case: for all three regimes
of $A_0=-1,0,1$ TeV, one has $1\% \lesssim \Delta m_{\tilde
  \ell}/m_{\tilde \ell}\ (\tilde e_{_L} , \tilde \mu_{_L}) \lesssim
10\%$. Remarkably, one can have $\Delta m_{\tilde
  \ell}/m_{\tilde \ell}\ (\tilde e_{_L} , \tilde \mu_{_L}) \sim 5\%$, in
agreement with current bounds on BR($\mu \to e \gamma$). Concerning 
the amount of LFV inducing the $\mu \to e \gamma$ transitions, one
finds that, similar to what occurs in the degenerate case, the largest
BRs are associated with $M_{N_{2}}$  close to its maximal allowed
value (i.e. $\sim M_{N_{3}}$, leading to degenerate
heavy and next-to-heavy triplets) and minimal values of $M_{1/2}$. 
While the latter leads to a lighter spectrum, the former allows to
enhance the $(Y^{\nu \dagger} L Y^\nu)_{21}$ contributions
proportional to  $M_{N_{2}}$, which are not suppressed by the
smallness of $\theta_{13}$.
However, it is important to notice that the same does not occur
regarding BR($\tau \to \mu \gamma$), which is maximal for both minimal
values of $M_{1/2}$ and $M_{N_{2}}$ (now degenerate with the lightest
triplet). 
For fixed values of $M_{N_{1,3}}$,
while the flavour violating entries responsible for $\mu \to
e \gamma$ transitions and other decays involving the first lepton family
(i.e. $(\Delta m^2_L)_{_{12}}$ and $(\Delta m^2_L)_{_{13}}$)
increase with increasing $M_{N_{2}}$, 
$(\Delta m^2_L)_{_{23}}$ - which induces
BR($\tau \to \mu \gamma$) - remains approximately constant: in fact it
actually decreases by a small factor, since the 
contributions proportional to  $M_{N_{2}}$ have the opposite sign of
those associated to $M_{N_{3}}$.

Finally, and to conclude our numerical study, we have considered 
deviations from the
$R=1$ limit, i.e., allowing for additional mixings in the seesaw
mediators. Non-vanishing angles $\theta_{i}$ lead to
larger $Y^\nu_{ij}$, with implications for LFV observables: as
expected (and aside from eventual accidental cancelations), 
there is a large enhancement of the contributions to
low-energy LFV observables, as well as an increase of the mass
splittings. More concretely, this would displace
the cyan regions in Figs.~\ref{fig:BR:deltaMS:emu} and 
\ref{fig:NDmasses:BR:deltaMS:emu} towards larger values of BR($\mu \to
e \gamma$) - potentially excluded by current bounds - and towards
slightly larger $\tilde e_{_L} - \tilde \mu_{_L}$  mass differences. 
Notice that when compared to the type I SUSY seesaw, the effects of 
$R\neq 1$ are somewhat less important, since due to the much 
narrower interval of the seesaw scale (which is also heavier), 
perturbativity of $Y^\nu$ effectively constraints the values of $\theta_{i}$.
Concerning the impact of these variations on the SUSY spectrum (RGE
induced), we have verified that deviations from 
$R=1$ have no effect on the gaugino and squark spectra.

\bigskip
To summarise,
let us re-emphasise that should the $\chi_2^0 \to \chi_1^0 \ \ell \ell$
decay chain be reconstructed at the LHC, a type III SUSY seesaw will
be manifest in both low- and high-energy LFV observables, which will lie within
the sensitivity of present/future experimental facilities.  
In other words, finding regions in the  type III SUSY seesaw parameter
space where the $\chi_2^0 \to \chi_1^0 \ \ell \ell$ is present,
without observing neither $\mu \to e \gamma$ transitions at MEG, nor
$\Delta m_{\tilde \ell}/m_{\tilde \ell}\ (\tilde e_{_L} ,
\tilde \mu_{_L})$ at the LHC, is almost impossible.

\section{Conclusions}\label{sec:conclusions}

Although it is a very appealing hypothesis to explain the origin of neutrino
masses and mixings, the seesaw mechanism is in general very hard to
probe directly. 
When embedded into a larger framework (as for instance SUSY models),
where new states are active between the seesaw scale and the
electroweak one, the seesaw mechanism can give rise to many distinct signatures,
depending on the nature of the mediators: 
scalar or fermionic (gauge singlets or triplets). 
In this study we  considered a supersymmetric type III seesaw where, 
in order to preserve gauge coupling unification, the
additional states are embedded into complete SU(5) representations.  
The many experimental
constraints (LEP, LHC, low-energy experiments)   
strongly reduce the available parameter space of the model, so that
one expects very characteristic signals (SUSY spectrum and  
charged LFV,  both at low-energies and at the LHC), which offer the
possibility of falsifying the model.    
Using the correlation between the different LFV observables
(inherent from the assumption that the seesaw provides the only source
of flavour violation in the model), we have focused our analysis on
the interplay between low-energy radiative decays (e.g. $\mu\to e
\gamma$) and potential LFV signatures appearing in association with
the $\chi_2^0\to \tilde \ell \ell$ cascade decays at the LHC, such as
flavoured slepton mass splittings, $\Delta m_{\tilde  \ell}/m_{\tilde
  \ell}\ (\tilde e_{_L} , \tilde \mu_{_L})$.  

\medskip
Firstly, requiring that the spectrum allows for the reconstruction of
slepton masses from the $\chi^0_2$ cascade decay chains  
(and assuming a $\chi^0_1$ LSP), 
the type III SUSY seesaw leads to scenarios where a heavy SUSY
spectrum (e.g. $m_{\tilde q} \sim 2$ TeV) is in general favoured.  
Although viable dark matter scenarios are in general very hard to
accommodate in the type III seesaw, we have nevertheless verified  
that one can still find small regions in the parameter space where the
$\chi^0_1$ has the correct relic density. Such scenarios  
typically arise in association with the low $m_0$ regime. Concerning
dark matter, it is important to recall that other candidates might be
present and have a relic density in agreement with WMAP bounds, as
could be the case for gravitinos. However, this issue clearly
lies beyond the scope of the present work. 

\medskip  
Assuming that a type III seesaw is indeed the only source of LFV, and
given the extremely constrained parameter space, one finds that
 the corresponding slepton mass splittings will always lie around the
 \% level, and are thus 
within the expected sensitivity of the LHC. 
A hierarchical  fermionic triplet spectrum further boosts the expected
mass splittings: one is led to a regime where, even in the
conservative limit  of $R=1$, one has
$1\% \lesssim \Delta m_{\tilde  \ell}/m_{\tilde \ell}\ (\tilde e_{_L},
\tilde \mu_{_L}) \lesssim 
5\%$, in agreement with current bounds on charged LFV.   
Furthermore, these mass
splittings correspond to values of BR($\mu \to e \gamma$) well within
the expected sensitivity of MEG (or, in very limiting cases, within
PRISM/PRIME sensitivity for CR($\mu-e$, Ti)).  

In the more general case of 
an increased mixing involving the triplet sector (i.e.
$R\neq1$), there is an enhancement of the contributions to
low-energy  LFV observables, as well as a small increase in the slepton mass
splittings, without further impact on the remaining SUSY spectrum. 

\bigskip
Unlike what occurs for a type I SUSY seesaw, the very constrained
range for the type III seesaw scale strongly tightens the predictions
for LFV:  
the expected flavoured mass splittings are indeed well within the
sensitivity range of the LHC, while at the same time low-energy scale
LFV must unavoidably lie within  the present and future sensitivity of
either MEG or PRISM/PRIME (observation of a $\tau \to \mu \gamma$
signal at SuperB will be much more challenging). If supersymmetry is
discovered at the LHC, and a type III seesaw is at the origin of
flavour mixing in the lepton sector, then this model can be easily 
falsified  in the near future.

\section{Acknowledgements}
This work has been done partly under the ANR project CPV-LFV-LHC {NT09-508531}. 
The work of  A. J. R. F.  has been supported by {\it Funda\c c\~ao
  para a Ci\^encia e a 
Tecnologia} through the fellowship SFRH/BD/64666/2009. 
A. J. R. F. and J. C. R. also acknowledge the financial support from
the EU Network grant UNILHC PITN-GA-2009-237920 and from {\it
Funda\c{c}\~ao para a Ci\^encia e a Tecnologia} grants CFTP-FCT UNIT
777, CERN/FP/83503/2008 and PTDC/FIS/102120/2008.

\end{document}